\documentclass [lettersize, twocolumn] {IEEEtran}
\usepackage{algorithm}
\usepackage{algorithmic}
\usepackage{array}
\usepackage{adjustbox}
\usepackage[caption=false,font=normalsize,labelfont=sf,textfont=sf]{subfig}
\usepackage{textcomp}
\usepackage{stfloats}
\usepackage{url}
\usepackage{verbatim}
\usepackage{graphicx}
\usepackage{cite}
\usepackage{longtable}
\usepackage{lipsum} 
\usepackage{amsmath,amssymb,amsfonts}
\usepackage{xcolor}
\usepackage{multirow}
\usepackage{subfig}
\usepackage{makecell}
\usepackage{enumitem}
\usepackage{derivative}
\usepackage{balance}
\usepackage{siunitx}

\usepackage[utf8]{inputenc}
\usepackage{float}
\usepackage{afterpage}
\usepackage{tikz}
\usepackage{diagbox}

\newcolumntype{P}[1]{>{\centering\arraybackslash}p{#1}}

\begin{document}
\def\UrlBreaks{\do\/\do-}
\title{Multimodal Instruction Disassembly with Covariate Shift Adaptation and Real-time Implementation}

\author{\IEEEauthorblockN{Yunkai Bai\textsuperscript{*},
Jungmin Park\textsuperscript{+},
Domenic Forte\textsuperscript{*}}

\IEEEauthorblockA{Department of ECE,
University of Florida, Gainesville, Florida, USA\textsuperscript{*}\\
Lucid Motors, Newark, CA, USA\textsuperscript{+} \\
Email: baiyunkai1995@outlook.com,
jungminpark@lucidmotors.com,
dforte@ece.ufl.edu}}


\markboth{}%
{Shell \MakeLowercase{\textit{et al. }}: A Sample Article Using IEEEtran.cls for IEEE Journals}

\IEEEpubid{}

\maketitle

\begin{abstract}
Side-channel based instruction disassembly has been proposed as a low-cost and non-invasive approach for security applications such as IP infringement detection, code flow analysis, malware detection, and reconstructing unknown code from obsolete systems.
However, existing approaches to side-channel based disassembly rely on setups to collect and process side-channel traces that make them impractical for real-time applications. 
In addition, they rely on fixed classifiers that cannot adapt to statistical deviations in side-channels caused by different operating environments.
In this article, we advance the state of the art in side-channel based disassembly in multiple ways.
First, we introduce a new miniature platform, RASCv3, that can simultaneously collect power and EM measurements from a target device and subsequently process them for instruction disassembly in real time. 
Second, we devise a new approach to combine and select features from power and EM traces using information theory that improves classification accuracy and avoids the curse of dimensionality.
Third, we explore covariate shift adjustment techniques that further improve accuracy over time and in response to statistical changes.
The proposed methodology is demonstrated on six benchmarks, and the recognition rates of offline and real-time instruction disassemblers are compared for single- and multi-modal cases with a variety of classifiers and over time. Since the proposed approach is only applied to an 8-bit Arduino UNO, we also discuss challenges of extending to more complex targets.
\end{abstract}


\section{Introduction}

\IEEEPARstart{S}{ide-}channels are most well-known for enabling non-invasive recovery of keys from cryptographic algorithms and of other assets used within microarchitectures.
However, side-channel leakage can also be used for defensive purposes. For example, reverse engineering original source code from side-channel measurements has a wide variety of applications, such as the detection of intellectual property (IP) infringement~\cite{park2018power}. Side-channel leakage could also be helpful in the code leakage detection\cite{eisenbarth2010building,shelton2021rosita++,mccann2017towards,kiaei2022soc,de2022armistice,sehatbakhsh2020emsim,buhan2022sok}. In some cases, users may already know the code but are not familiar with its functionality. By analyzing side channels, such users can non-invasively determine which parts of the code are more often executed and find ways to improve  software performance.



An interesting and more recent defensive application of side-channel analysis is anomaly detection in critical systems~\cite{forte2024nowhere}. 
Here, anomalies can be caused by software and hardware faults, exploitation of vulnerabilities, or malware. The latter two can be linked to supply chain attacks on third-party and open-source software, which are increasingly compromising critical infrastructure, as seen in high-profile incidents like the SolarWinds attack~\cite{alkhadra2021solar}. 
Traditional malware and vulnerability detection methods, including signature-based and behavior-based approaches~\cite{souri2018state}, often fall short against sophisticated threats due to scalability issues and more. 
To mitigate these risks, the use of Software Bills of Materials (SBOMs) is recommended for documenting all software components in a product or system, enabling better software composition analysis and proactive vulnerability management~\cite{brief2021executive}. 
However, even SBOMs have limitations with respect to unknown vulnerabilities. 
Further, even if patches are available to address a threat identified in an SBOM, the availability requirement of critical systems might make them challenging to deploy. 
Side-channel based malware detection can address unknown vulnerabilities by \textit{detecting unavoidable changes in a systems' side channels} caused by the malware\cite{forte2024nowhere}. 
In addition, doing so in real-time provides a new ability to monitor critical systems with known malware/vulnerabilities, allowing one to take actions to prevent serious consequences \textit{only when they are being exploited}.

Side-channel based anomaly detection can be categorized as coarse- or fine-grained. Coarse-grained detection with side-channels can be applied to identify the presence or absence of higher level program structures~\cite{nazari2017eddie}. For example, some measure the loop time or perform spectrum analysis of the side-channel traces and compare them with benign traces~\cite{nazari2017eddie}. Clark et al~\cite{clark2013wattsupdoc} proposed WattsUpDoc for detecting malware on embedded medical devices. They collect power traces and use supervised machine learning (ML) with traces of both normal and abnormal activity. Their algorithms could achieve 94\% detection rate towards known
attacks and at least 85\% accuracy for unknown malware. Xiao et al.\cite{xiao2017nipad} proposed NIPAD to detect abnormal activities in a programmable logic controller (PLC), and achieve 90.33\% recognition rate towards the malicious modifications of the original
program. Kim et. al\cite{kim2008detecting} introduced a framework to monitor, detect, and analyze unknown energy-depletion threats. They pre-store a signature and compare it to the power leakage in the testing phase. A 99\% true-positive rate and less than 5\% false-negative rate are obtained in classifying malware from an HP iPAQ running inside a Windows Mobile OS.
Although coarse-grained methods could achieve high accuracy against malware, they still struggle to detect small changes in code, from high latency, and from the inability to monitor systems in real time~\cite{khan2019malware}.

Fine-grained malware detection capture side-channel measurements and use them to recover instructions or bit encodings. The former is often referred to as side-channel based instruction disassembly. 
Traditional disassemblers were developed for reverse engineering instructions by implementing algorithms to decode machine code into instructions\cite{paleari2010n}\cite{schwarz2002disassembly}\cite{nanda2006bird}. 
Similar to traditional disassemblers, side-channel based disassemblers (SCDs)\cite{strobel2015scandalee}\cite{cristiani2020bit}\cite{vaidyan2020instruction} often aim to infer a target system's instructions. 
However, the key difference is that side-channel disassemblers reverse engineer instructions from the side-channel EM/power leakage of the target device without modifying the hardware, rather than using machine code as traditional disassemblers do. 
It is important to emphasize that in the rest of this paper, the term ``disassembler" is used only to refer to side-channel disassemblers, not traditional disassemblers.

The first power SCD was introduced in 2010 by Eisenbarth et al.~\cite{eisenbarth2010building}. 
In~\cite{msgna2014precise}, Markantonakis et al. adopted the k-nearest neighbor method with Euclidean distance as its similarity metric to create their own power-based SCD for the ATmega163. 
In~\cite{park2018power}, Park et al. proposed a hierarchical classification approach that improved the scalability of power-based SCD. 
In~\cite{fendri2022deep}, Fendri et al. developed a deep learning framework to disassemble 36 instructions inside a RISC-V CPU.  
In addition to power traces, electromagnetic (EM) waves have also been explored for SCD more recently.
In 2015, Strobel et al.~\cite{strobel2015scandalee} disassembled instructions from the PIC microcontroller using EM measured using a self-made EM  taken at 20 different positions to achieve the best recognition rates to date. 
In~\cite{cristiani2020bit}, Cristiani et al. proposed a greedy algorithm to determine the best subset of EM  positions and were able to recover more than instruction opcodes. 
Vaidyan et al.~\cite{vaidyan2020instruction} adopted an upsampling method to track two target instructions using the EM frequency spectrum.

More details on all these SCD approaches and a comparison with the proposed approach can be found in Section~\ref{sec:rw}. 
Here, we highlight three major challenges with using the existing SCD methods for real-time defense and malware detection.
First, they adopt a traditional side-channel measurement system (e.g., oscilloscope, commercial EM , motorized stage, etc.) to collect traces as well as a PC to process traces, extract features, and classify them as instructions. 
While such setups are feasible in a lab environment, they \textit{cannot be used for in situ monitoring of real systems and operational technologies (OTs)}.
The equipment needed to perform disassembly could be much larger and more expensive than the target being monitored in the case of an IoT-based OT!
Second, instruction disassembly in real time is also more challenging because \textit{every instruction needs to be correctly recovered within limited time} and with just a single trace.
Thus, one cannot use the high-accuracy EM disassemblers discussed above because they require the program to be run multiple times and traces to be captured at a different  position during every run.
Finally, in practical applications, the stable circumstances experienced by a target embedded system in the lab during training often do not hold after deployment. 
For example, unavoidable restarts of embedded processors can create DC shifts in their side-channels~\cite{park2018power}. 
The existing classifiers do not account for changes in the input distribution (i.e., covariate shifts), \textit{lack the ability the generalize to new circumstances}, and will therefore have lower accuracy in the field than in training.


In this article, we systematically address these limitations. 
First, we build upon our remote access to side-channel (RASC) platform~\cite{bai2022rascv2} -- a miniature side-channel measurement, processing, and classification system. 
The third version discussed in this article has higher sampling speed and resolution, more memory, and a larger FPGA than our previous one, making it suitable for real-time instruction disassembly. 
Second, rather than only relying on a single modality (power \textit{or} EM) to perform instruction disassembly, we take advantage of RASC's ability to simultaneously measure \textit{both} power and EM channels. 
Complementary and redundant information from the two channels is used to improve classification accuracy. 
Even though there are a few works that combine power and EM channels to improve the efficiency of side-channel attacks~\cite{standaert2008using,souissi2012towards,bai2023dual}, this is the first article to do so for instruction disassembly to the best of our knowledge. 
Last but not least, we overcome non-stationary conditions of the target's environment and associated side-channel probability distributions by adjusting the traces and classifier parameters at run time. 
This enables our dissassembler to generalize and improve its accuracy over time.

Our main contributions are summarized as follows:

\begin{itemize}

\item A mathematical analysis is provided to quantify the gains made by combining EM and power channels and to identify their optimal combination. The suitability for real-time instruction disassembly on the resource-limited RASCv3 is further improved through lightweight dimensionality reduction where mutual information is utilized to select the best time indices of the dual channel traces.


\item Measurements are collected from an Arduino UNO target device using offline and real-time setups. The former is a traditional side-channel setup with an oscilloscope and commercial EM probe, while the latter collects power and EM using RASCv3. The efficiency of multiple feature sets and classifiers are compared by recognition rate vs. number of features. The results show that dual-mode (power \textit{and} EM) approaches significantly outperform single channels (power only or EM only) across multiple classifiers (LDA, QDA, and MLP). Further, the proposed lightweight feature reduction approach gives comparable or better performance than the more heavyweight PCA.

\item Testing is also performed on six real benchmarks using both offline and real-time modes. In the offline mode, a PC is used to perform disassembly. In the latter, feature combination and selection (parameterized offline), QDA classifiers (trained offline), and a portion of the self-adjustment algorithm are implemented on RASCv3 to perform disassembly in real time. The Arduino UNO is turned on and off throughout the testing in order to intentionally cause covariate shifts. Although the real-time mode disassembly accuracy is lower than offline mode (80\% vs. 90\% on average), both exhibit improvements over time from self-adjustment.

\item The challenges of performing real-time SCD on more complex targets are discussed and it is empirically shown that the sampling rate of the side-channel measurements should approximately 40$\times$ the target's clock frequency for maximum accuracy.

\item The RASCv3 design files, bill of materials, etc. have been uploaded to GitHub~\cite{RASCv3PCB} for use by the research community.

\end{itemize}

The remainder of this paper is organized as follows: Section~\ref{sec:back} provides background information on classification, mutual information, and feature selection. 
Section~\ref{sec:disassemblyflow} details the high-level SCD methodology and workflow used in this article. Section~\ref{sec:meth} discusses the integration of dual-channel (power and EM) features. Section~\ref{sec:meth: real} explains the real-time implementation of the proposed methodology, including covariate shift minimization. Section~\ref{sec:result} presents the experimental setup, results, and discussion. Challenges and directions for extending real-time SCD to more complex targets is discussed in Section~\ref{sec:complex_targets}. Section~\ref{sec:rw} compares related work with the proposed methods and Section~\ref{sec:conclusion} concludes the paper.

\section{Background} \label{sec:back}
This section introduces background on classification, mutual information, entropy, and feature reduction/selection. 

\subsection{Classification Task Formulation}\label{sec:back:task}

A general classification problem involves assigning input features to predefined categories or classes, and the goal is to develop a model that can accurately predict the appropriate class for new, unseen data points. Typically, a function $\Phi$ maps input features ($x$) to $n$ possible classes 
\begin{equation}\label{formula:classifier}
{c= \Phi(x;\Theta)}, 
\end{equation}
where $c \in \{c_{1},c_{2}, \hdots, c_{n}\}$ and $\Theta$ represents the classifier and its parameters. In this article, $x$ consists of side-channel traces, and $c$ is a processor instruction.

Based on the number of classes, classification could be divided into binary-class ($n=2$) and multi-class ($n>2$) types. There are multiple ways to achieve multi-class classification from binary classification. For example, in a ``one vs. one'' approach (OvO), a binary classifier is created for every pair of classes, and the predicted class is the one with the most votes from all the binary classifiers. On the other hand, the ``one vs all'' approach (OvA) creates a binary classifier for each class, treating it as the positive class and the rest as the negative class. During prediction, the class with the highest confidence of all binary classifiers will be selected. Since OvO is more expensive than OvA, we use OvA in this paper. However, we also experiment with true multi-class classifiers that rely on deep learning to specify one out of $n$ classes given the input features.

\subsection{Classifiers of Interest}\label{sec:back:classifierofinterest}
In this article, we consider three potential classifiers for instruction disassembly, many of which have been popular in the related work: 
\begin{enumerate}
\item\textbf{Linear Discriminant Analysis (LDA)} relies on the assumption that each
class is drawn from a multivariate Gaussian distribution with a class-specific mean row vector and a covariance matrix.
For LDA, all classes have the same covariance matrix. 
 
\item\textbf{Quadratic Discriminant Analysis (QDA)} shares similar algorithms as LDA. However, the discriminant analysis results from class $c_i$'s specific mean row vector ($\vec{\mu}_{c_i}$) and covariance matrix ($\boldsymbol{\Sigma}_{c_i}$)~\cite{james2013introduction}. The QDA classifier adopts the discriminant function to calculate the posterior probability of an observation belonging to a specific class. The discriminant function of unknown sample $\vec{x}'$ belonging to class $c_i$ is
\begin{align}\label{formula:qda:decision}
\delta_{i} = & -\frac{1}{2}{(\vec{x}'-\vec{\mu}_{c_i})}^{T}\boldsymbol{\Sigma}_{c_i}^{-1}(\vec{x}'-\vec{\mu}_{c_i}) \nonumber\\ 
& + \log(p_{c_i}) - \frac{1}{2}\log(|\boldsymbol{\Sigma}_{c_i}|)
\end{align}
where $p_{c}$ is the probability of class $c$. The final classification output of  $\vec{x'}$ is decided by the index of the highest decision value:
 \begin{equation}\label{formula:qda:output}
c_i = \mathop{\arg\max}_{i}\{\delta_{i}\}.
\end{equation}

\item The \textbf{Multilayer Perceptron (MLP)} is a fully connected class of feedforward artificial neural network (ANN) and consists of at least three layers of nodes: an input layer, a hidden layer, and an output layer. Besides the input layer, each layer inside the MLP uses a nonlinear activation function, which defines the output of that node given an input or set of inputs, including ReLU and Sigmoid. The MLP algorithm used in this article is from~\cite{MLP}.
\end{enumerate}

QDA and LDA are traditional statistical classifiers while MLP is a deep learning method. Generally speaking, QDA and LDA classifiers produce high classification rates for low-dimensional feature input, especially in cases of limited training data availability. MLP is more useful in domains with large and high-dimensional data~\cite{janiesch2021machine}.

\subsection{Mutual Information}

Mutual information is a quantity that measures the mutual dependence between two random variables\cite{cover1999elements}. In the area of disassembly, mutual information can be adopted to describe the correlation between the sampling points in side-channel traces (i.e., features) and instructions. In this article, we will later utilize it to quantify the relationship between features derived from both power and EM signals and the instructions.

Assuming any feature is a discrete random variable $X$, and the instruction is a discrete random variable denoted as $C$, the mutual information between $X$ and $C$ is given by
\begin{equation}\label{formula:1}
I(X,C)= \sum_{x \in X}^{}\sum_{c \in C}^{} P_{X,C}(x,c)
\log \left( \frac{P_{X,C}(x,c)}{P_{X}(x)P_{C}(c)} \right ), 
\end{equation}
where $P_{X} (x)$ and $P_{C} (c)$ are the marginal probability mass functions of random variables $X$ and $C$. Here, $x$ and $c$ are possible values taken by $X$ and $C$. $P_{X,C} (x, c)$ is the joint distribution and the marginal distributions are $P_{X}$ and $P_{C}$.

Based on information theory, the mutual information (Equation \eqref{formula:1}) could also be expressed in terms of entropy~\cite{cover1999elements}:
\begin{equation}
{I(X,C)= h(X)-h(X|C)}. \label{formula:2}
\end{equation}
Here, $h(X)$ stands for the entropy of the target feature $X$, and $h(X|C)$ means the entropy of the target feature $X$ when given class label $C$. In this article, we normalize all features into the range $[0,1]$. For this case, the range of $X$ is limited to $[0,1]$ and its probability density function $p$ obeys the Gaussian distribution with the mean $\mu$ and the standard deviation $\sigma$. The entropy of $X$ could therefore be written as 
\begin{align}\label{formula:4}
h(X)&=-\int_{0}^{1}{p(x)\log(p(x))dx} \nonumber \\
&=-\int_{0}^{1}{p(x) \left [ -\frac{1}{2}\log(2\pi\sigma^2)-\frac{x^2}{2\sigma^2}\log(e) \right]dx} \nonumber \\
&=\frac{1}{2}\log(2\pi\sigma^2)+\frac{\sigma^2}{2\sigma^2}\log(e) \nonumber \\
&=\frac{1}{2}\log(2{\pi}e\sigma^2).
\end{align}
Equation~\eqref{formula:2} can be rewritten using Equation~\eqref{formula:4}
to express mutual information with the standard deviation of the whole dataset and the standard deviation of each class:
\begin{equation}
I=\frac{1}{2}(\log(2{\pi}e\sigma^2)- \sum_{i=1}^{n}p_{C_{i}}\log(2{\pi}e\sigma_{c_{i}}^2), \label{formula:8}
\end{equation}
where $\sigma$ is the standard deviation of whole dataset, $\sigma_{c_{i}}$ denotes the standard deviation of the data from class $i$, and $p_{c_{i}}$ represents the probability of class $i$ in a whole dataset. In this article, we refer to mutual information expressed using Equation~\eqref{formula:8} as the Gaussian approximation format. 

\subsection{Feature Selection Methods}\label{sec:rt:fs}

In this paper, we compare four feature reduction/selection methods: mRMR, PCA, Gini, and Filter.

\textbf{mRMR (minimum redundancy max relevance)}~\cite{peng2005feature} maximizes the relevance between the selected features and the target class while minimizing the redundancy among the selected features. 
This approach helps in selecting a subset of features that are highly relevant to the target variable and minimally redundant with each other, resulting in a more efficient classifier. 
In this article, after showing its robustness in preliminary experiments, we adopt mRMR for real-time SCD in the remainder on the paper. More details of mRMR will be provided in Section~\ref{sec:back:mRMR}.

\textbf{PCA (principle component analysis)}\cite{PCA} reduces the dimensionality of large data sets by linearly combining all its variables into a smaller set that still contains most of the information (specifically, variance) of the larger set. 
An advantage of PCA is that it can be applied in an unsupervised way to the input data; that is, class labels are not needed to use it for feature reduction. 
We refer the reader to~\cite{PCA} for more details.

\textbf{Gini index}\cite{8361371} is a measure of impurity or purity of a dataset. It quantifies the likelihood of a randomly chosen element being incorrectly classified if it was randomly labeled according to the distribution of labels in the dataset. The Gini index for a set of items with \( n \) classes is calculated according to the following equation,
\begin{equation}\label{fs:Gini}
\text{Gini} = 1 - \sum_{i=1}^n p_i^2
\end{equation}
where \( p_i \) is the probability of an item being classified in the $i$th class.

\textbf{Filter} selection\cite{Duch2006} is a type of feature selection technique that relies on the intrinsic properties of the data. It evaluates each feature independently of the classifier and ranks  them based on a statistical measure. The key formula of the filter method in this article is mutual information, which is listed in Equation~\eqref{formula:1}.

\section{High-level disassembly methodology}\label{sec:disassemblyflow}

\begin{figure}
  \centering
  \hfill
{\includegraphics[width=0.48\textwidth]{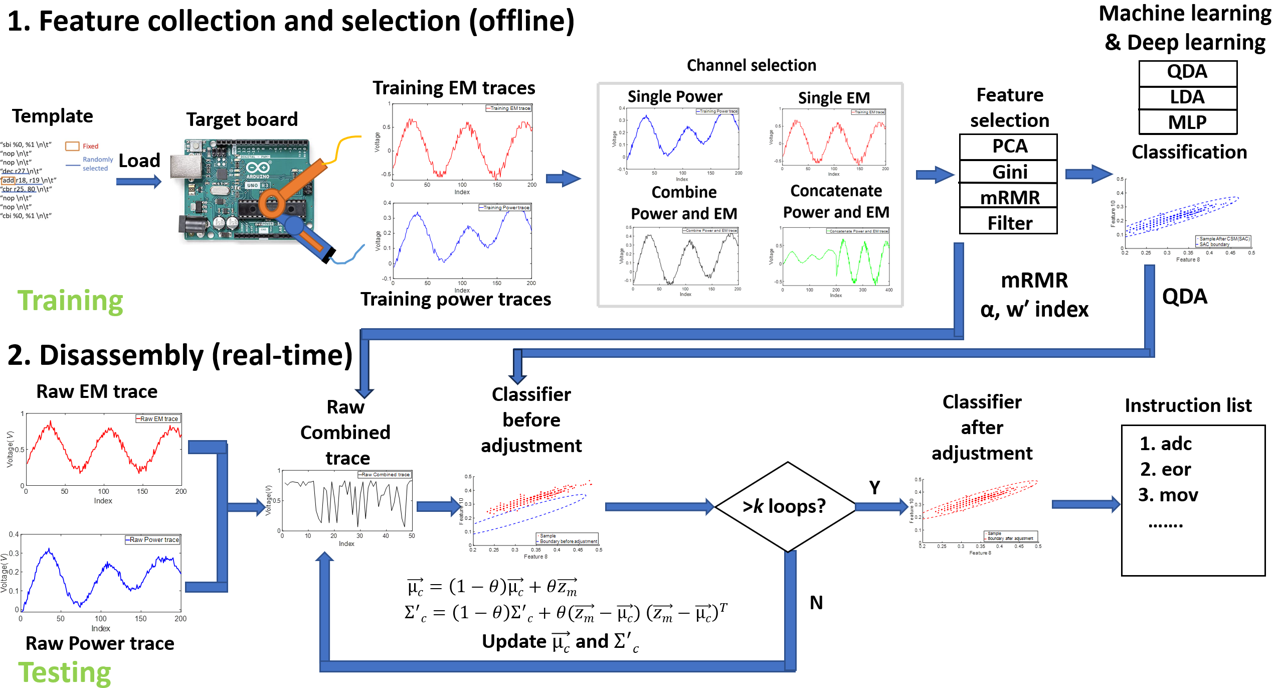}}
\hfill
  \caption{High-level diagram of the proposed methodology. 
  (a) Feature Selection Phase: The goal in this phase is to identify the optimal combination of feature selection methods (including coefficients and indices) and classification algorithms. Initially, EM and power traces (red and blue, respectively) are collected from the target board. Different channel configurations are then evaluated, including power only, EM only, and linearly combined EM and power using mutual information. Five feature selection methods are considered: PCA, mRMR, Gini index and Filter selection. Following feature selection, deep learning and machine learning classifiers (LDA, QDA, and MLP) are employed to disassemble the instructions. By the end of this phase, the combination of EM/Power traces, mRMR feature selection, and a QDA classifier is selected for real-time disassembly;
  (b) Real-Time Disassembly Phase: This phase aims to perform SCD in real-time.
  The dual-channel traces are collected and combined using the mRMR method. To reduce the impact of environmental changes on results, the classifiers are fine-tuned using covariate shift minimization. Once the adjustments are completed, the subsequent combined test traces are fed into QDA classifiers to obtain disassembly outcomes.}  \label{fig:methodology_highlevel}
\end{figure}

This section briefly introduces the high-level methodology for offline and real-time SCD. The overall workflow and design space is presented in Figure~\ref{fig:methodology_highlevel}.


\subsection{Measurement Setups}\label{sec:disassemblyflow:setup}

In this paper, we use offline and real-time measurement setups and compare the results.  The offline setup uses an MDO3102 oscilloscope to sample power and EM from the target -- an Arduino UNO which is a microcontroller based on the ATmega328P. The real-time setup uses a custom platform that we designed called RASC. For both setups, power is measured from a shunt resistor. In the offline setup, EM is measured using a commercial probe (LF1 Set from LANGER EMV Technik\cite{EMV}). The real-time setup uses antennas printed on the RASC PCB or printed using a Dimatix printer. More details on RASC and the real-time measurement setups can be found in Section~\ref{sec:RASC}.

These measurements are collected in an offline training mode where the most important samples in a trace are identified and used to train classifiers. In the real-time mode, measurements are collected, processed, and classified using RASC. A high-level overview of feature selection and classification are given in the next subsections.

\subsection{Feature Collection and Selection} \label{sec:disassemblyflow:fs}

\begin{figure}[t]
\centering
\subfloat[]{
\includegraphics[width=0.235\textwidth]{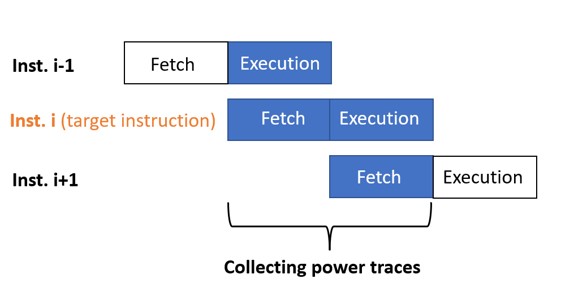}}
\hfill
\subfloat[]{
\includegraphics[width=0.235\textwidth]{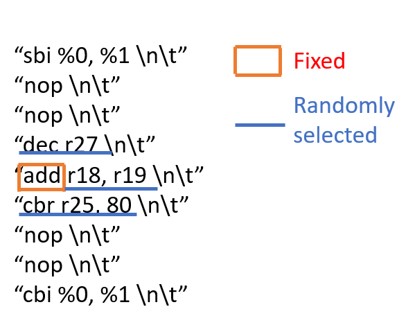}}
\caption{(a) Instructions in the pipeline during clock cycles $i-1$, $i$, and $i+1$; (b) Example template used to train a classifier for the \texttt{add} instruction.}
\label{fig:example}
\end{figure}

At the start of the feature collection, we create a training template for all instructions based on their similarity in the microcontroller architecture. For Arduino UNO, there exist two pipeline stages. One fetches the instructions, and another one executes the instructions. The power and EM traces of the target (current) instruction could be affected by previous and subsequent instructions which are also in the pipeline. Thus, the training template for Arduino mainly consists of NOPs, a random instruction, the target instruction, another random instruction, and NOPs as shown in Figure~\ref{fig:example}.

 \begin{table*}[t]\setlength{\tabcolsep}{4pt}
 \scriptsize
\centering
 \caption{Grouping of AVR instructions for hierarchical classification.}   \label{table:group}
\begin{tabular}{|c|c|c|c|c|c|c|c|c|}
\hline
 & Group 1 & Group 2 & Group 3  & Group 4 & Group 5 & Group 6 & Group 7 & Group 8\\ 
 \hline
Instructions &\makecell{ADC, ADD,\\ AND CP,\\
CPC, CPSE,\\  EOR, MOV, \\OR, SBC, \\SUB, MOVW}  & \makecell{ADIW, ANDI, \\CBR, CPI, \\LDI, ORI,\\  SBCI, SBIW, \\SBR, SUBI} 
& \makecell{ASR, CLR, \\COM, DEC, \\INC, LSL,\\ LSR, NEG, \\ROL, ROR, \\SER, SWAP,\\ TST } 
&  \makecell{BRCC, BRCS, BREQ,\\ BREG, BRHC, BRHS,\\ BRLO, BRLT, BRMI,\\ BRNE, BRPL, BRSH,\\ BRTC, BRTS, BRVC,\\ BRVS, CALL, JMP,\\ RCALL, RJMP} 

& \makecell{LD,\\LDD,\\LDS} 

&\makecell{LPM,\\ ELPM}   

&\makecell{CLC, CLN, CLS,\\ CLT, CLV, CLZ,\\ SEC, SEH, SEI,\\ SEN, SES, SET,\\ SEV, SEZ} 

& \makecell{BCLR, BLD, \\BRBC, BRBS,\\ BSET, BSTCBI,\\ SBI, SBIC,\\ SBIS, SBIC,\\ SBRS }  \\ 
\hline
Operands & \makecell{Rd, Rr} & \makecell{Rd, K} &\makecell{Rd}  & \makecell{k} & \makecell{Rd, k\\Rd, (-)X(+)\\Rd, (-)Y(+(q))\\Rd, (-)Z(+(q))}& \makecell{Rd, Z(+)} & \makecell{} & \makecell{Rr(Rd),b\\A, b\\s,k\\s} \\   
\hline
\makecell{Number of Instructions} & \makecell{12} & \makecell{ 10}  &\makecell{13 }   & \makecell{20} & \makecell{3} &\makecell{2}  & \makecell{14}  &\makecell{12}   \\    
 \hline
Description & Arith.  & Arith.,Data.   &Bits,Arith. & Bran.  &Data.  &Data.  &Bits. &Bran.,Bit. \\
 \hline

\end{tabular}
\end{table*}

Three different types of traces are investigated: Power channel only, EM channel only, and Power and EM combined, which are denoted by $P$, $EM$, and $Z$, respectively. We refer to the first 2 as single-channel and the latter as dual-channel. We empirically select a $w$-size feature window in the trace for the classification. Thus, from this point forward, the index of each trace will be described in the range [1, $w$]. Features in the traces are selected using four different feature selection approaches (PCA, mRMR, Gini and Filter). For the dual-channel case, we normalize each power and EM feature into the range [0, 1]. Then, the EM and power features are linearly combined and a subset of them are selected according to the approach described in Section~\ref{sec:meth}. That is, we determine the optimal combination coefficients ($\alpha_k^*$) for each feature index $k$ to maximize the mutual information for the $k$th combined (or dual channel) feature, i.e., $Z_k = \alpha_{k}^* P_k + (1-\alpha_{k}^*)EM_k$, $1\leq k \leq w$. The feature indices and their combination coefficients are saved (for later use in real-time) and used to train classifiers.

At the end of the feature selection phase, a hierarchical classification with inter-group classifiers and within-group classifiers is utilized. In this article, we adopt the same grouping for instructions as in~\cite{park2018power} (see Table~\ref{table:group}). The hierarchical classification significantly reduces computational complexity in the case of the classification of large classes (86 in our case) significantly. For example, if OvO binary classifiers are exploited, 3655 classifiers should be trained. On the other hand, using the hierarchical OvA strategy, 568 classifiers are necessary
for all instructions: 568 =$\binom {8} {2}$+$\binom {12} {2}$+$\binom {10} {2}$+$\binom {13} {2}$+$\binom {20} {2}$+$\binom {3} {2}$+$\binom {2} {2}$+$\binom {14} {2}$+$\binom {12} {2}$. At this step, we train all the classifiers using an instance of the target.

Finally, as illustrated in Figure~\ref{fig:methodology_highlevel}, we consider three different classifiers (MLP, QDA, and LDA). The offline experiment results in Section~\ref{sec:offline} indicate that the mRMR feature selection combined with the QDA classifier is the most suitable choice for real-time disassembly. More details can be found in Section~\ref{sec:offline}.

\subsection{Real-time disassembly}\label{sec:disassemblyflow:rt}
In the real-time SCD phase, we collect EM and power test traces from a real running code and generate combined dual channel traces ($Z$) for $w^*$ features using the feature selection approach described above. We perform inter-group and within-group classification. The classifier is then fine-tuned based on the
classification outcomes to minimize covariate shift. The real-time implementations of feature selection, hierarchical classification, and self-adjustment process are explained in Section~\ref{sec:meth: self-enhance} in detail. After that, the adjusted classifier repeats to disassemble an entire program code. Any significant deviations from the expected code can be attributed to malware (e.g., see~\cite{bai2022real}).

\section{Dual Channel Combination \& Feature Selection}\label{sec:meth}

\begin{table}[t]\setlength{\tabcolsep}{4pt}
 \scriptsize
\centering
  \caption{Table of notation, variables, and abbreviations}
  \begin{tabular}{{|P{0.5in}|P{2.5in}|}}
  \hline
       \textbf{Notation} & \textbf{Explanation} \\
        \hline          \texttt{{}$\vec{x}$ }& A lowercase letter with an arrow above it indicates a vector\\  
    \hline
    \texttt{{}$\vec{x}^T$ }& Transpose of a vector\\  
    \hline
        \texttt{{}$\boldsymbol{X}$ }& An upper case letter or symbol in bold indicates a matrix\\
    \hline
        \texttt{{}$X, x$ }& Upper and lower case letter pairs stand for discrete random variables and their possible values\\
    \hline
        \texttt{{}$\mathbb{X}$}& Upper case blackboard bold letters represent a set\\
    \hline
         \texttt{{}$\mu_x$},$\sigma_x$ & Mean and standard deviation of $x$\\
     \hline
          \texttt{{}$\boldsymbol{\Sigma}_x$}& Covariance matrix of $\vec{x}$\\
    \hline

    \hline
     \textbf{Variable} & \textbf{Explanation} \\
     \hline
    \texttt{{}$I$ } &Mutual information \\
     \hline
    \texttt{{}$c_i$ } & Label associated with the $i$th class (instruction)\\
     \hline
    \texttt{{}$p_{c_i}$ } & Probably of $i$th class (instruction)\\
     \hline
    $\sigma$ & Standard deviation\\

    \hline
    \texttt{{}$k$ } & Number of traces to adjust classifier\\
     \hline
    
    \texttt{{}$\alpha$ } &Linear combination coefficient  \\
    \hline
    \texttt{{}$Z$}& Linear combination of power and EM channels at a time index according to $\alpha$\\
    \hline

    \texttt{{}$\dot{o}_{Z}$}& Approximation of mutual information
for combined feature $Z$\\

    \hline
        \texttt{{}$f(\alpha)$}& $\dot{o}_{Z}^2$  \\
    \hline
        \texttt{{}$v(\alpha)$}& Key factor of $\odv{f(\alpha)}{\alpha}$  \\
     \hline
    \texttt{{}$\alpha^*$ }& Analytically calculated optimal combination coefficient, i.e., zero point of function $v(\alpha)$ \\  
    \hline
    $w'$ & Number of selected feature during the process of mRMR\\
    \hline
     \texttt{{}$w$ }& Feature window (vector) size of a trace used to classify any instruction \\  
    \hline   
             \texttt{{}$\theta$ }& Coefficient in QDA classifier self-adjustment\\  
    \hline

\texttt{{}$\mu_{c}$,$\boldsymbol{\Sigma}_{c}$, $b_{c}$}& Mean, covariance, and bias parameters for class label $c$ in QDA classifier\\    \hline
    
   
\texttt{{}$G(X,Y)$ }&Relevance of a feature $X$ for class label $Y$\\  
    \hline    
\texttt{{}$R(X,Y)$ }&Redundancy of features $X$ and $Y$ \\  
    \hline

    $\mathbb{S}'$, $\mathbb{S}$ &Selected and non-selected feature sets\\
    \hline
    \texttt{{}$M_i$ }&Mutual information difference when moving the $i$th feature from set $\mathbb{S}$ to $\mathbb{S}'$\\  
    \hline
   \texttt{{}$H_1$,$H_2$ }&Random variable in channel 1, Random variable in channel 2\\
        \hline
        \hline
    \textbf{Abbreviated Term}  & \multirow{2}{*}{\textbf{Explanation}} \\
    \hline
     OvA & One vs. all classification\\  
    \hline
     OvO & One vs. one classification\\  
    \hline
        MID & Mutual Information Distance\\  
    \hline    
    \hline    
    mRMR & minimum redundancy max relevance\\
    \hline    
    PCA & Principle component analysis\\
    \hline
    Gini & Gini index feature selection\\
    \hline
    Filter & Filter feature selection\\
    \hline%

  \end{tabular}

  \label{table:notation}
\end{table}

This section discusses the proposed methodology for combining EM and power features. The EM and power side channels are assumed to be captured simultaneously and consist of the same number of samples. Section~\ref{sec:meth:pre} introduces the preconditions of the combination. Section~\ref{sec:meth: MIGuass} derives the optimal coefficient for linearly combining power and EM features. Finally, Section~\ref{sec:back:mRMR} describes how we select a subset of combined EM/power features from traces for dimensionality reduction.

The notation and parameters used throughout the article are listed in Table~\ref{table:notation} to assist the reader.

\subsection{Preconditions}\label{sec:meth:pre} 

Before deriving the condition for a better combination than the single channel, we first emphasize three preconditions for disassembly and explain them in sequence. A target feature represented as a discrete random variable is denoted as $X$. Then, we assume

\begin{enumerate}
\item The probability of the selected feature of the EM/power trace ($p(x)$)  obeys a Gaussian distribution.
\item Given class $c_i$ of $n$ classes, the probability of the selected feature of the EM/power trace ($p(x|c_i)$) obeys the Gaussian distribution.
\item The probability of all classes are equal to each other, i.e., $p(c_i=1)=p(c_i=2)= \hdots p(c_i=n)$.
\end{enumerate}

For the first precondition, the collected EM/power traces from the same group of instructions share similarities in functionality and operands but act differently in opcodes. Besides, we normalize all EM/power traces into the range [0,1]. Thus, the normalized EM/power features naturally obey the Gaussian distribution. In this case, for calculating mutual information between power/EM features and the target instruction, we can use the Gaussian approximation  (Equation~\eqref{formula:8}).

Note that we adopt the hierarchical instruction classification approach used in~\cite{park2018power} where similar instructions are grouped together, and two types of classifiers are used. The first type of classifier predicts the instruction's group from the trace (i.e., \textit{inter-group classification}). From there, the second type of classifier predicts the exact instruction from within the group (i.e., \textit{within group classification}). For multi-class classifier schemes, this makes SCD more easily realized on a resource-constrained device like RASC. 

For the second precondition, class $c_i$ refers to either the $i$th group number for inter-group classification or the $i$th type of instruction for within-group classification. The size of the class is decided by how many groups are inside the inter-group classification and how many instructions are in the within-group classification.
The given class $c_i$ means the target instruction is fixed during training. In this article, the target instructions also remain the same in the example template. However, we randomly vary the instructions surrounding the target, the destination register, and processed values for the same target instruction. Thus, it is plausible that we may still take the Gaussian approximation for the second precondition given the randomness that this introduces.

The third precondition assumes the chance of each instruction in the within-group classification and each group in the inter-group classification are the same. In the training, we collect the same amount of traces of each instruction for within-group classification and the same number of traces of each group for inter-group classification. Thus, it is reasonable to let the probability of all classes be equal to each other. Note, however, that when performing disassembly on real benchmarks, this pre-condition needs to be relaxed since the prior probabilities of each instruction are not necessarily equal. This can be improved upon in future work, e.g., by incorporating a Hidden Markov Model (HMM)
 like in~\cite{eisenbarth2010building}.

Under these three preconditions and Equation~\eqref{formula:8}, the mutual information between target feature $X$ and the class label $C$ could be transformed into 
\begin{equation}
I=\frac{1}{2}\log \left (\frac{\sigma^2}{\prod_{i=1}^{n}\sigma_{c_{i}}^\frac{2}{n}} \right ) \label{formula:9}
\end{equation}
where $n$ stands for the total number of classes (value differs per group), $\sigma$ is the standard deviation of the target feature, and $\sigma_{c_{i}}$ is the target feature's standard deviation for class $i$. 

\begin{figure}[t]
  \centering
\subfloat[]{
  \includegraphics[width=1.55in]{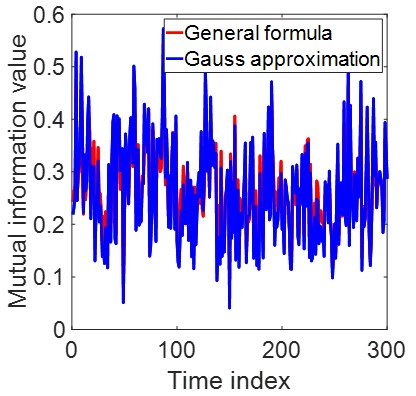}} 
\subfloat[]{
  \includegraphics[width=1.55in]{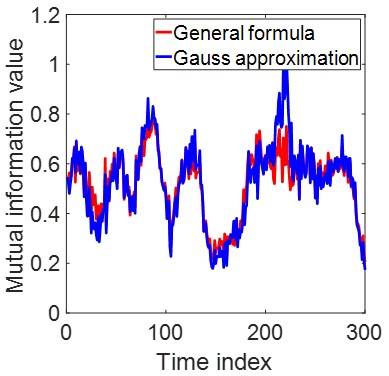}}
        \caption{Mutual information between (a) EM and class label; (b) power and class label.}
        \label{fig:MIEM}
\end{figure}

Figure~\ref{fig:MIEM} shows the mutual information of EM and power traces with class labels using two different approaches: ``Gaussian approximation'' uses Equation~\eqref{formula:9} and the ``General formula'' uses Equation~\eqref{formula:1}. The calculated mutual information value between the first group instruction traces and the instruction label under the two calculation methods is almost identical. In other words, the three preconditions of Gaussian distribution seem to hold up well in practice.

\subsection{Optimal Combination of Features}\label{sec:meth: MIGuass}

\begin{figure}[t]
  \centering
\includegraphics[width=0.48\textwidth]{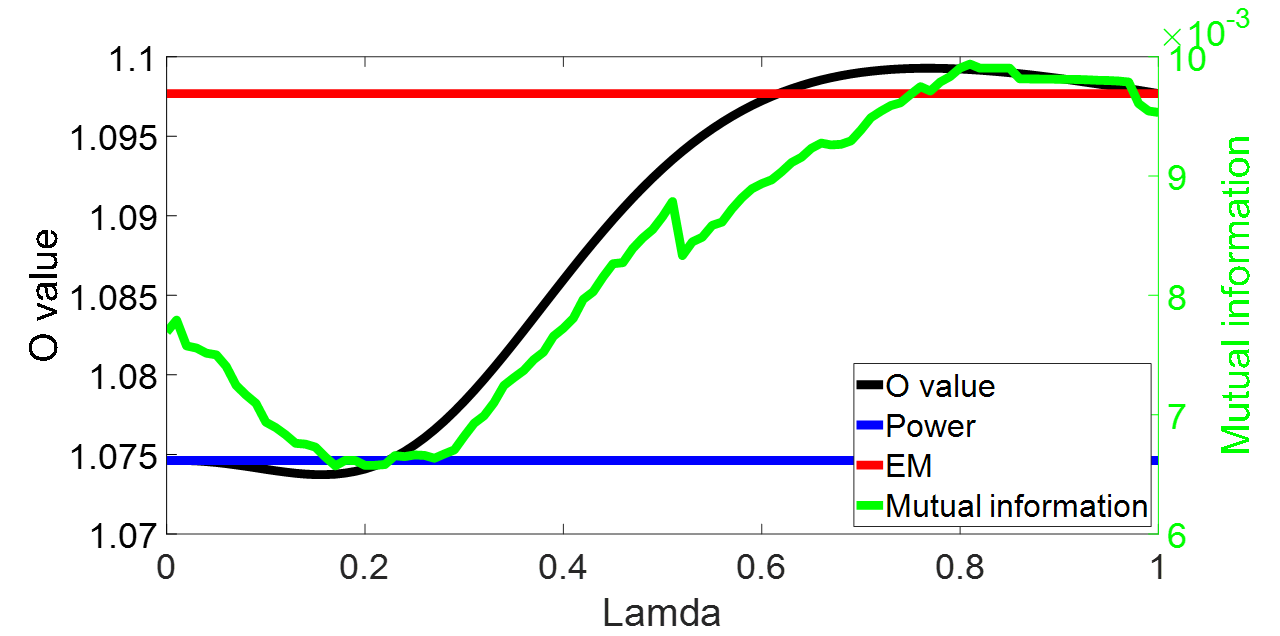} 
  
        \caption {Combined 1st feature's linear coefficient $\alpha$ vs.  mutual information and $\dot{o}$.}
       \label{fig:mutualcomparison}
\end{figure}

Here, we succinctly present our approach for combining features from the same time index of EM and power traces. More detailed derivations can be found in the conference version of this article~\cite{bai2023dual}.
Based on information theory~\cite{cover1999elements}, mutual information ($I$) is greater than or equal to 0. Thus, we can get the following inequality from Equation~\eqref{formula:10}:
\begin{equation}\label{formula:10}
\frac{\sigma^2}{\prod_{\forall i}^{}\sigma_{c_{i}}^\frac{2}{n}} \geq 1.
\end{equation}
Further, we use the symbol $\dot{o}$ to describe the $n$th power of Equation~\eqref{formula:10}:
\begin{equation}\label{formula:11}
\dot{o}=\frac{\sigma^n}{\prod_{ i=1}^{n}\sigma_{c_{i}}}.
\end{equation}
In Equation~\eqref{formula:12}, the features from the first and second channel at the same time index are denoted by random variables $H_1$ and $H_2$, and both are normalized to the range $[0,1]$. Here, we adopt $\alpha$ to describe the linear combination coefficient and $Z$ to describe the combined feature:
\begin{equation}\label{formula:12}
{Z=\alpha H_1+(1-\alpha) H_2}. 
\end{equation}
Note that $Z \in [0,1]$ as long as $\alpha \in [0,1]$. We can derive Equation~\eqref{formula:16} for $Z$, 

\begin{equation}\label{formula:16}
\dot{o}_{Z}=\frac{\sigma_{H_1}^n}{ \left(\frac{\sigma_{H_1}}{\sigma_{Z}} \right)^n\prod_{i=1}^{n}\sigma_{Z_{C_i}}}.
\end{equation}
Finally, the condition that supports the improvements gained from a combined feature is
\begin{equation}\label{formula:17}
\left(\frac{\sigma_{H_1}}{\sigma_{Z}} \right)^n\prod_{\forall i}^{}\sigma_{Z_{C_i}} \geq \prod_{\forall i}^{}\sigma_{H_{1_{C_i}}}.
\end{equation}
Empirically, Equation~\eqref{formula:17} is satisfied by some choice of $\alpha$. Take Figure~\ref{fig:mutualcomparison} as an example -- the maximum of mutual information of the 1st combined feature of group 1 instructions (shown in green) is around $\alpha=0.8$. 

For determining the value of $\alpha$ that makes the highest mutual information of the combined channel, we should find $\alpha^* = {\arg\max}_{\alpha}(\dot{o}_Z)$. Further, since $\dot{o}_Z \geq 0$, this is equivalent to ${\arg\max}_{\alpha}(\dot{o}_Z^2)$. The analytical derivation is provided as follows:
\begin{equation}\label{formula:20}
\dot{o}_Z^2= f(\alpha)=\prod_{\forall i}^{}f_{i}(\alpha)
\end{equation}
and
\begin{equation}\label{formula:19}
f(\alpha)= \frac{{\sigma_Z}^{2n}}{\prod_{\forall i}^{}\sigma_{Z_{c_{i}}}^2}
=\prod_{\forall i}^{}\frac{{\sigma_Z}^2}{\sigma_{Z_{c_{i}}}^2}.
\end{equation}
$\alpha^*$ can be calculated by setting the first derivative of $f(\alpha)=0$. This derivative can be written as 
\begin{equation}\label{formula:21}
\odv{f(\alpha)}{\alpha}
=\left(\prod_{\forall i}^{}f_{i}(\alpha) \right) \left(\frac{2\alpha(1-\alpha)}{\alpha^2\sigma_{H_1}^2+(1-\alpha^2)\sigma_{H_2}^2}\right) v(\alpha)
\end{equation}
where
\begin{equation}\label{formula:22}
v(\alpha)=\sum_{\forall i}\frac{\sigma_{H_1}^2\sigma_{H_{2_{C_i}}}^2
-\sigma_{H_2}^2\sigma_{H_{1_{C_i}}}^2}{\alpha^2\sigma_{H_{1_{C_i}}}^2+(1-\alpha)^2\sigma_{H_{2_{C_i}}}^2}.
\end{equation}
Examining the middle term in Equation~\eqref{formula:21} yields two trivial zero points that occur at $\alpha=0$ and $\alpha=1$. 
Thus, we define 
\begin{equation}
    \alpha^* = \alpha |_{v(\alpha)=0}
\end{equation}
Empirically, for the first feature's coefficient, the calculated maximum point using this formulation (shown in black) is 0.77 in Figure~\ref{fig:mutualcomparison}, which is very close to the empirical result of 0.8. Also shown in Figure~\ref{fig:mutualcomparison} are the mutual information for the first feature corresponding to power only (blue) and EM only (red). The combined feature using the optimal coefficient has larger mutual information.

\subsection{Feature Selection}\label{sec:back:mRMR}

In this article, we adopt mRMR (minimum redundancy max relevance)~\cite{peng2005feature} as the primary method for feature selection. mRMR considers both ``redundancy'' and ``relevance'' so that each newly selected feature is not only highly correlated with the class label but also avoids redundancy with previously selected features. This section formally introduces the notions of relevance and redundancy and then explains how the mRMR's Mutual Information Difference (MID) criteria can be used to select the ``best'' combined EM-power features.

Relevance describes the degree to which a feature is closely related to a class label. Obviously, higher relevance between the selected features and the class label is beneficial to classification. The relevance ($G$) between combined feature $i$ ($Z_{i}$) and class label $C$ is given as
\begin{equation}\label{formula:MIrelevance}
G(Z_{i},C)= \sum_{x \in Z_{i}}^{}\sum_{c \in C}^{} P_{Z_{i},C}(x,c)
\log \left( \frac{P_{Z_{i},C}(x,c)}{P_{Z_{i}}(x)P_{C}(c)} \right ), 
\end{equation}
where $P_{Z_{i}}(x)$ and $P_{C} (c)$ are the marginal probability mass functions of $Z_{i}$ and $C$. $P_{Z_{i},C} (x, c)$ is the joint distribution and the marginal distributions are $P_{Z_{i}}$ and $P_{C}$.

\begin{figure*}[t]%
\centering
 \subfloat[]{
\includegraphics[width=0.3\textwidth]{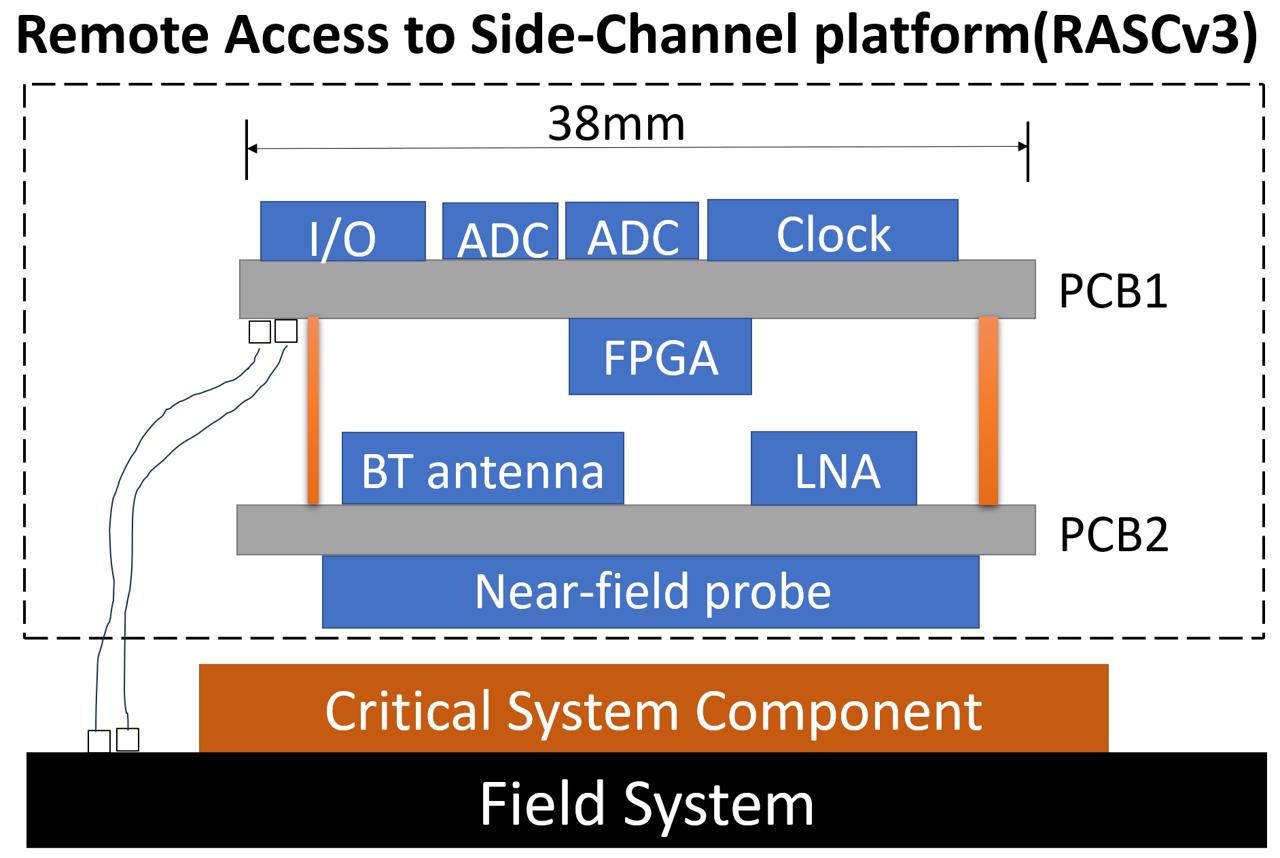}}
\hfill
 \subfloat[]{
\includegraphics[width=0.29\textwidth]{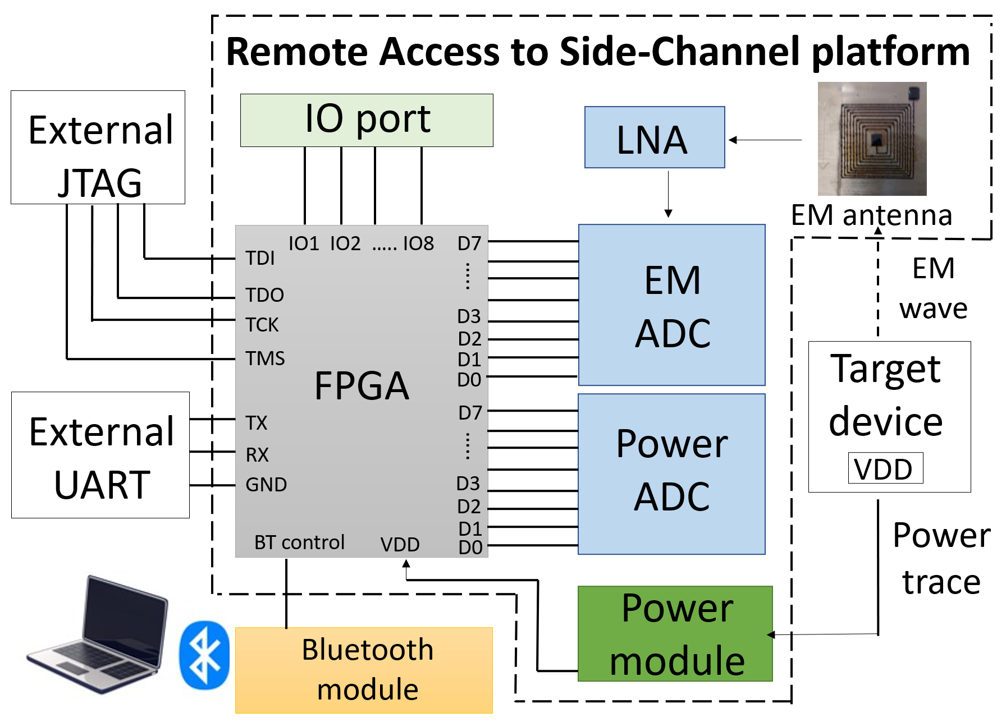}}
\hfill
\subfloat[]{
  \includegraphics[width=0.29\textwidth]{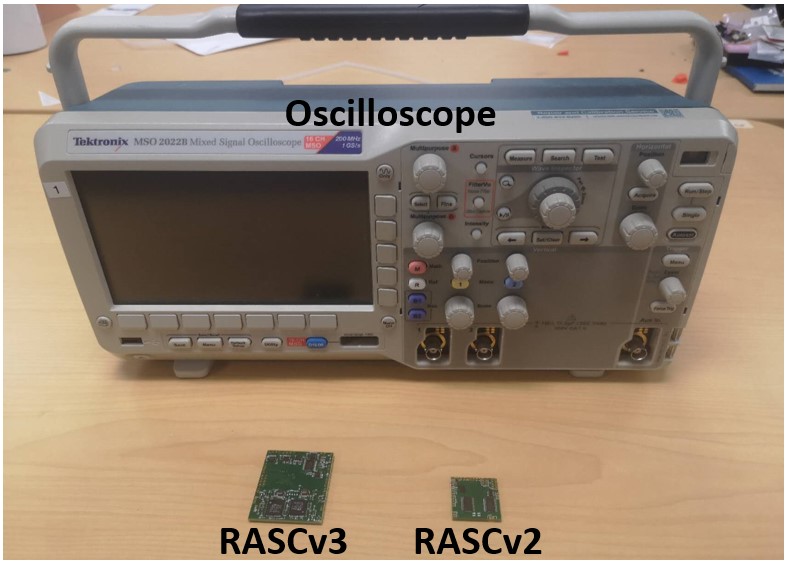}}
\caption{(a) Depiction of RASC in the field.} RASCv3 is a 30 mm by 30 mm PCB board with chips for signal processing, communication, and memory. RASC also measures the target's power and EM traces; (b) Depiction of the RASC schematic. RASC may be powered by the $V_\textrm{dd}$ pin of the target. The traces are digitalized by ADC chips on RASC. JTAG and UART modules are used to program the FPGA and transmit data; (c) Size comparison between an oscilloscope (Tektronix MDO 3102), RASCv3, and RASCv2.
\label{fig: RASCV1V2structure}
\end{figure*}

Redundancy refers to repetitive information between features that does not add further value to the selected feature set. Evidently, higher redundancy between a newly added feature and the selected feature could make classification less efficient. In this article, the redundancy ($R$) between combined $i$th and $j$th feature set $Z_{i}$ and $Z_{j}$ is calculated as follows: 
\begin{equation}\label{formula:MIredanduncy}
R(Z_{i},Z_{j})= \sum_{x \in Z_i}^{}\sum_{y \in Z_j}^{} P_{Z_{i},Z_{j}}(x,y)
\log \left( \frac{P_{Z_{i},Z_{j}}(x,y)}{P_{Z_{i}}(x)P_{Z_{j}}(y)} \right ), 
\end{equation}
where $P_{Z_{i}}(x)$ and $P_{Z_{j}}(y)$ are the marginal probability mass functions of features $Z_{i}$ and $Z_{j}$,  $P_{Z_{i},Z_{j}}(x,y)$ is the joint distribution, and the marginal distributions are $P_{Z_{i}}(x)$ and $P_{Z_{j}}(y)$.

Suppose the combined dataset owns $w$ features in total and there have already been $w'$ features selected. The selected and non-selected feature sets are $\mathbb{S'}=\{s'_{1}, s'_{2},\hdots,s'_{w'}\}$ and $\mathbb{S}=\{s_{w'+1}, s_{w'+2},\hdots,s_{w}\}$. When using mRMR to select the next feature, one needs to compare these two sets using the Mutual Information Difference (MID) criteria, which tries to balance the relevance and redundancy of the features used.
The MID for feature index $i$ from $\mathbb{S}$ is calculated as 
\begin{equation}\label{formula:MTD}
M_{i}=G(s_{i},C)-\frac{1}{w'}\sum_{s'_j \in \mathbb{S}'}^{}{R(s_{i},s'_{j})} 
\end{equation}
Equation~\eqref{formula:MTD} is used to calculate the MID value for each feature in  $\mathbb{S}$. The new selected feature $s'_{w'+1}$ is the one with the largest MID, i.e., 
\begin{equation}\label{formula:select}
s'_{w'+1}=\arg\max_{w'+1 \leq i \leq w} \{M_{i}\}.
\end{equation}
Note that for the initially selected feature ($w'=1$), there is no selected feature index and only redundancy is considered for selection. 

\section{Real-time SCD Implementation}\label{sec:meth: real}

This section discusses all aspects of our implementation of real-time SCD elements, including RASC (platform for capturing, processing and classifying traces), power and EM sensing, adaptive classifiers, and feature selection. Preliminary results which guide our choice of sensing methods, classifiers, and feature selection are also discussed.

\subsection{Remote Access to a Side-Channels (RASC)}\label{sec:RASC}

RASC is an external monitor which minimizes the traditional side-channel system into two tiny PCB boards. The first version of RASC (RASCv1) was introduced in~\cite{stern2019rasc} while the second version of RASC (RASCv2) appeared in~\cite{bai2022real}. In this article, we design a third version of RASC (RASCv3) and successfully load codes into it to disassemble instructions in real-time mode. 

Figure~\ref{fig: RASCV1V2structure}(a-b) depicts the RASCv3 deployment scenario and schematic. RASCv3 contains two ADCs\cite{LT2242} for digitizing power and EM traces simultaneously, an external Bluetooth module like HC-05 for remote communication, and a Xilinx Artix-7 100t FPGA~\cite{FPGA} for data processing. We also implement 8 I/O ports on the board and connect them to an external UART module to transmit trace data for communication with a PC. When it is working, as depicted in Figure~\ref{fig: RASCV1V2structure}(a), RASCv3 is attached to or arranged near the target and connects to its power supply. In this scenario, RASC\footnote{In this paper, we will often refer to RASCv3 as simply RASC. When it is ambiguous, we will use v2 and v3.} is not only powered but can also collect power traces of the target device in real time. If necessary, RASC could also connect to an external power module, such as a battery, to power it. The maximum transmission distance is over 30 meters, which is far enough to support remote monitoring and communicate any detected anomalies to a base station.

\begin{table}[t]\setlength{\tabcolsep}{4pt}
 \scriptsize
\centering
  \caption{Comparison between traditional side-channel analysis setup, commercial setups, and RASC.}
  \begin{tabular}{|c|c|c|c|c|}
  \hline
     \diagbox[]{\textbf{List}}{\textbf{Setup}} & \textbf{RASCv2} & \makecell{\textbf{ChipWhisperer} \\\textbf{Lite 32-Bit}}& \textbf{RASCv3}& \textbf{Oscilloscope}\\
     \hline
    \textbf{Cost} &\$150   & \$250 & \$400 & $\geq$\$12k \\
    \hline
    \textbf{Size}& 2.5$\times$2.5\si{\centi\meter}$^2$  & 12$\times$5.1\si{\centi\meter}$^2$& 3.8$\times$3.8\si{\centi\meter}$^2$ &  14.7$\times$50\si{\centi\meter}$^2$\\
    \hline
    \makecell{\textbf{Voltage} \\\textbf{range}}& [0V, 3V] & [-1V, 1V]& [-1V, 1V] & [-20V, 
 20V]\\
    \hline
    \makecell{\textbf{Sampling} \\\textbf{speed}}& 128MS/s & 105MS/s & 160MS/s & 5GS/s\\
    \hline
    \textbf{Resolution}& 8-bit& 10-bit& 12-bit  & 16-bit\\  
    \hline
    \makecell{Processing} \\\textbf{module}& \makecell{Spartan 3e\\FPGA}& \makecell{ATSAM3U2CA\\MCU}& \makecell{Artix-7\\FPGA}  & None\\  
    \hline
    \makecell{\textbf{Remote} \\\textbf{communication}}& Yes & No & Yes & No\\
            \hline
  \end{tabular}
  \label{Comparision}
\end{table}

\subsection{RASC Specifications vs. Disassembly}

RASC’s specifications that impact its disassembly performance include~\cite{forte2024nowhere}:

1) \textbf{Sampling rate} which refers to the number of side-channel samples per unit time. Important features that distinguish instructions may be undersampled or missed entirely if the sampling rate is too low. 

2) \textbf{Sample resolution} refers to the number of bits per sample after ADC conversion. Lower sample resolution of power and EM side channels may result in a loss of information, potentially making it more challenging to capture subtle variations that could be indicative of class differences and or anomalies.

3) \textbf{Processing capabilities} affect RASC’s ability to handle large volumes of measurement data, sophisticated feature extraction and classification algorithms, and adaptation to nonstationary environments. Parallel processing in an ASIC or FPGA can significantly accelerate computations compared to an MCU. RASC’s specs should be chosen based on the complexity of the target chip (see Section~\ref{sec:complex_targets}) and the accuracy required for anomaly detection. The higher the sampling rate, sample resolution, and processing capabilities, the higher RASC’s cost and power requirements.

Table~\ref{Comparision} compares RASCv3 to a traditional oscilloscope, RASCv2, and the ChipWhisperer Lite 32-Bit board~\cite{Chipwhisperer} in terms of these hardware specs, size, cost, and other capabilities\footnote{The actual chips used in RASC can be found on our GitHub~\cite{RASCv3PCB}}. The Chipwhisperer Lite 32-Bit is a popular educational board designed by NewAE Technology Inc. for studying side-channel attacks. A traditional side-channel analysis system consists of an oscilloscope and an EM probe. For example, the oscilloscope used in this article is Tektronix MDO 3102, which costs over \$12,000. RASCv2 collects EM and power traces with two 8-bit ADCs under 128MS/s sampling speed and processes them with simple calculations. RASCv3 upgrades the second version to higher sampling speed (200MS/s at maximum) and higher sensitivity (12-bit ADC), and is more portable to connect a printed EM antenna. The cost of RASCv3 is fair and around \$400\footnote{Note that this cost should be considerably reduced when RASC is produced at higher volumes. Furthermore, our prototype contains additional test, debug, and communication features that might not be needed in practical applications. Removing them can further reduce its cost.}. RASCv3 is only 38mm $\times$ 38mm and much smaller than a typical oscilloscope. The tiny size of RASCv3 allows it to be easily arranged close to the target system. RASC can also communicate disassembly and malware detection results to security administrators.



\subsection{Power and EM Sensing}

RASC's ability to perform disassembly is also impacted by the methods used to physically measure power and EM. Power can be measured using shunt resistors in series with the power supply, differential probes, integrated power monitors, etc. The signal-to-noise ration (SNR) of an EM antenna can be affected by its size, structure, and material.


\begin{figure}
  \centering
  \begin{minipage}{0.2\textwidth} 
    \centering
    \includegraphics[height=6cm]{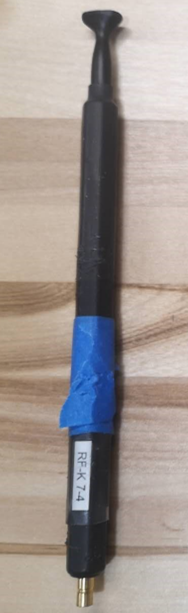} 
  \end{minipage}
  \hspace{0.1cm}
  \begin{minipage}{0.2\textwidth} 
    \centering
    \begin{minipage}{\textwidth} 
      \centering
      \includegraphics[width=1.8cm]{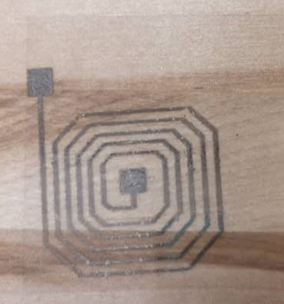}
      \vspace{0.3cm}
    \end{minipage}  
    \\
    \begin{minipage}{\textwidth} 
      \centering
      \includegraphics[width=1.8cm]{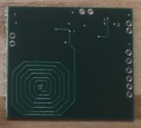}  
      \vspace{0.3cm}
    \end{minipage}  
    \\
    \begin{minipage}{\textwidth} 
      \centering
      \includegraphics[width=1.8cm]{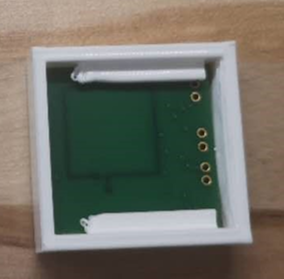} 
    \end{minipage}
  \end{minipage}
  \caption{Near-field antenna comparison. Left is the commercial EM , Right top is the printable antenna for RASCv3 board 3. The right middle is the near-field antenna on RASCv3 board 2. The right bottom is the internal antenna inside RASCv2 board 2.}
  \label{antennacomparison}
\end{figure}

To collect power leakage, a 10-ohm shunt resistor was soldered between the GND pin of the ATmega328P chip and the GND pin on the Arduino UNO board. The power traces were captured by connecting the two terminals of the shunt resistor to a nearby SCA detection tool, such as the MDO3102 oscilloscope and RASC in offline and real time setups, respectively. For capturing EM traces, we tested four antennas: a commercial EM probe (Figure~\ref{antennacomparison} left), a near-field antenna on the back of RASCv3 (Figure~\ref{antennacomparison} right middle), a near-field antenna inside RASCv2 (Figure~\ref{antennacomparison} right bottom), and a printable antenna (Figure~\ref{antennacomparison} right top). The commercial EM probe is the RF-K74 from LANGER EMV Technik\cite{EMV}. The near-field antenna inside RASCv2 was first introduced in~\cite{bai2022rascv2}. The design of the near-field antenna on the bottom layer of RASCv3 was improved based on lessons learned from the RASCv2 antenna, whose SNR was affected by the connection between internal vias~\cite{bai2022rascv2}. To address this, we placed the antenna on the bottom layer of RASCv3. To avoid potential short-circuit issues, as shown by the white frame in Figure~\ref{antennacomparison} (right bottom), a holder was designed to support RASC. Additionally, the printable near-field antenna shown in Figure~\ref{antennacomparison} (right top) could be attached to RASC. This antenna was printed using the DMP-2850 from Dimatix and NPS-L ink from Iwatani Corporation, composed primarily of silver and isohexadecane. By printing the antenna, we can easily iterate its design, enabling RASCv3 to adapt to more challenging scenarios.

\begin{figure}
  \centering    \includegraphics[width=0.45\textwidth]{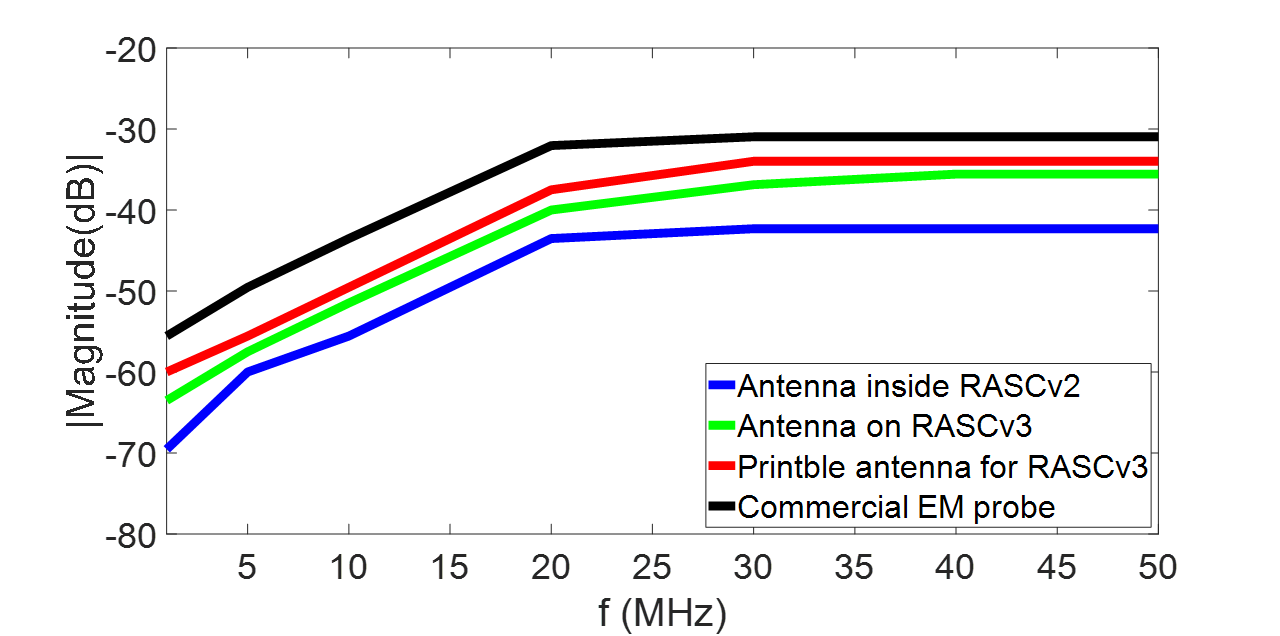}
     \caption {Near-field antenna magnitude comparison. }
     \label{magnitude_comparison}
\end{figure}

The size of the near-field antenna in RASCv2 (Figure~\ref{antennacomparison} right bottom) was 2cm$\times$2cm, and 4 turns of the near-field antenna were separated into 4 internal layers of PCB2. The received EM traces of RASCv2 were effective in earlier SCA experiments~\cite{bai2022rascv2}. However, the transmission efficiency was affected by the internal connection between PCB layers, and the structure of the internal layers increased RASCv2's cost. Thus, in RASCv3, we designed a 4 turn one-layer antenna on a PCB and printed a similar one using the DMP-2850. 

In Figure~\ref{magnitude_comparison}, we present the magnitude versus frequency of those four antennas.
All antennas are set to the same position 2mm above the Arduino UNO GND port during the response magnitude testing. Eqn.~\eqref{formula:db} explains how we calculate the magnitude. 
\begin{equation}
M=20\log \left (\frac{V_{test}}{V_{ref}} \right ) \label{formula:db}
\end{equation}
$V_{ref}$ stands for the reference voltage of the signal source (3V in our experiment). $V_{test}$ denotes the response voltage of the antennas during testing. The EM response of the 4-turn internal antenna inside RASCv2 has the lowest magnitude among the options. Luckily, we used a 100$\times$ amplifier circuit on RASCv2 to augment the signal. The detected voltage from RASCv3 is higher than the internal antenna inside RASCv2. Compared with the antenna on the bottom of RASCv3, the printable antenna has 1.5$\times$ the circumference and 2$\times$ wider traces. For the commercial EM probe, it has the highest magnitude among all four antennas in testing. For the cost, the commercial EM probe is around \$800 while the internal antenna for RASCv2 is \$100. The printable antenna is the cheapest and it costs around \$1. Besides, our designed antennas (internal and printable) are more flexible to use. They can be situated on the target chip to work with either RASC or an oscilloscope. However, the commercial EM probe can only be matched with an oscilloscope and holder.

 \begin{figure}[t]
  \centering
\subfloat[]{
  \includegraphics[width=0.24\textwidth]{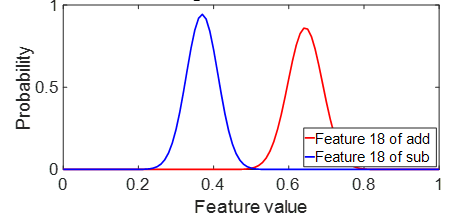}} 
       
\subfloat[]{
  \includegraphics[width=0.22\textwidth]{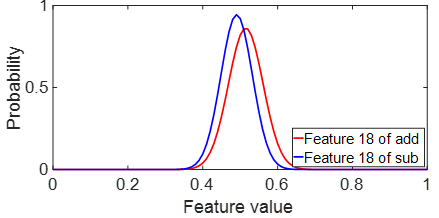}}
\subfloat[]{
  \includegraphics[width=0.24\textwidth]{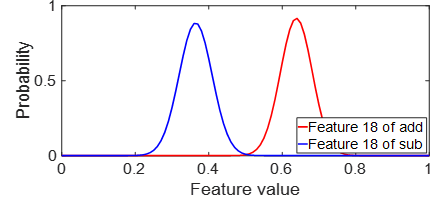}}
       \caption {Distribution comparison between feature 18 of instructions \texttt{add} and \texttt{sub} during (a) training; (b) testing before CSM; and (c) testing after CSM.} 
        \label{fig:distribution_comparison}
\end{figure}

\subsection{Self-enhancing Classifier}\label{sec:meth: self-enhance}


For solving the covariate shift issues, a self-enhancing step for the QDA classifier(s)~\cite{chen2013application} is needed. This self-enhancing step
works as a feedback process to the instruction classifier in real-time
applications. The self-enhancing algorithm adopts an incremental mode, and the parameters of classifiers are continuously adjusted to incoming data~\cite{chen2013application}. In the SCD task, the self-enhancing method could be combined with many traditional classifiers like LDA and QDA~\cite{chen2013application} and adjust their parameters in real time. Here, we describe it for QDA for simplicity. 

Before the QDA classifier adjustment phase, we have generated the QDA classifier and gotten all parameters, such as the original mean ($\vec{\mu}_{c_i}$) and covariance matrices ($\Sigma_{c_i}$) of class label $c_i$ (introduced in Section~\ref{sec:back:classifierofinterest}). Take combined trace segment $\vec{z_{m}}$ as an example. We first adopt the QDA classifiers before adjustment to determine its class label $c$, and set an update coefficient $\theta$. Note that $\theta$ could be decided by the ratio of the number of testing traces over the size of training traces~\cite{chen2013application}. Then, the adjusted mean value ($\vec{\mu}'_{c}$) and covariance matrix ($\boldsymbol{\Sigma}'_{c}$) for class $c$ are 
\begin{equation}\label{formula:SA1}
{\vec\mu'_{c} = (1 - \theta) {\vec\mu}_{c}+ \theta \vec{z}_{m}};
\end{equation}
\begin{equation}\label{formula:SA2}
{\boldsymbol{\Sigma}'_{c} = (1 - \theta) {\boldsymbol{\Sigma}_{c}} + \theta (\vec{z}_{m} - {\vec{\mu}}_{c}) (\vec{z}_{m} - \vec{\mu}_{c})^{T}}.
\end{equation}
With Equations~\eqref{formula:SA1} and \eqref{formula:SA2}, the mean and covariance matrices of class labels of classified incoming segments can be gradually adjusted.

Experimental data shows the impact of the adjustment. Figure~\ref{fig:distribution_comparison} compares the probability distributions for the 18th feature of instructions \texttt{add} and \texttt{sub}. The probability distribution of the same feature index for different instruction have different means in the training dataset (a), but are close to each other when collected in real-time testing (b). Covariate shift minimization (CSM), however, is capable of separating these means as shown in Figure~\ref{fig:distribution_comparison}(c).

\begin{figure}[t]
  \centering
\includegraphics[width=0.45\textwidth]{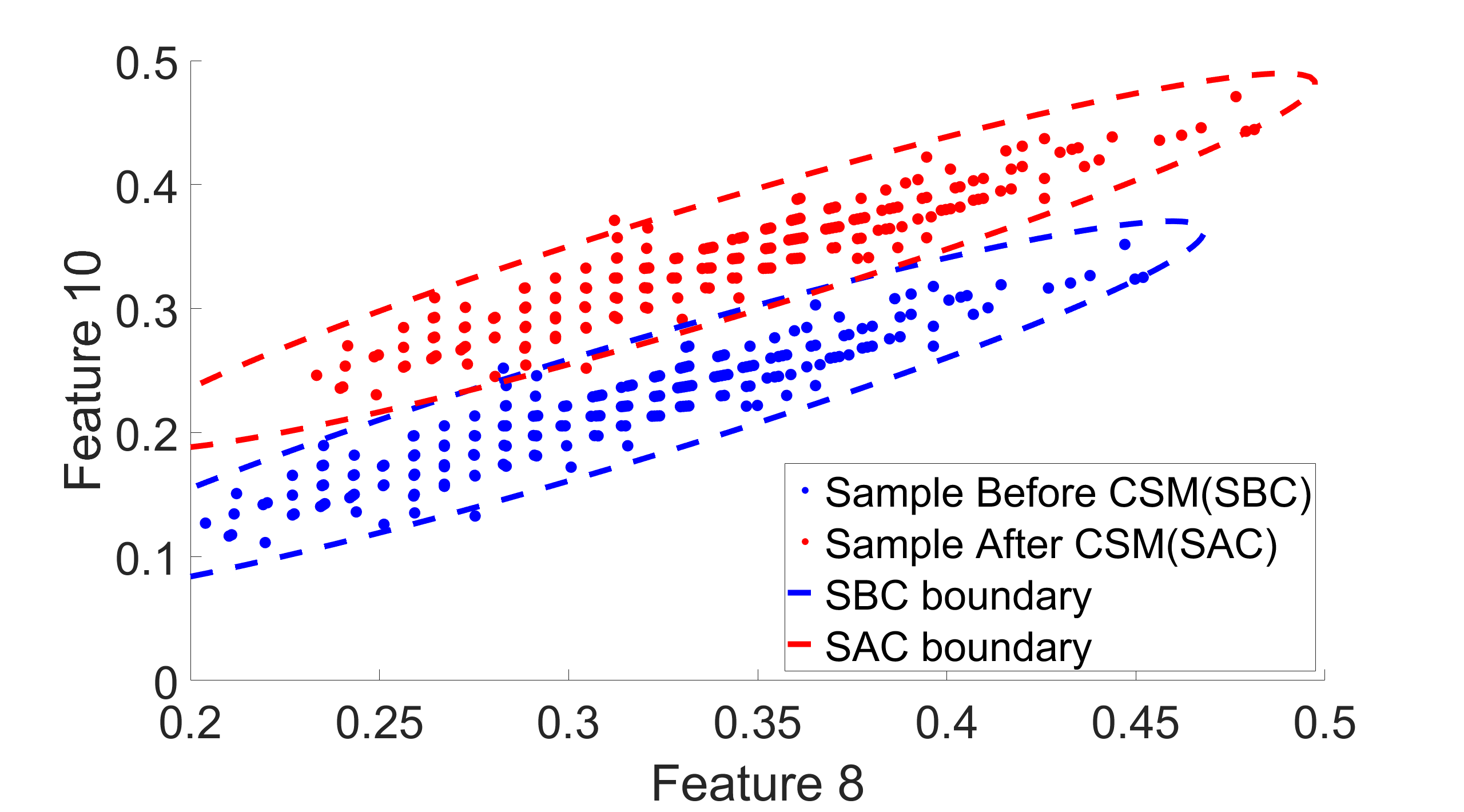} 
        \caption {Comparison of real-time samples of features 8 and 10 for instruction \texttt{add} before and after CSM without and with self-adjusted QDA boundaries.}
       \label{fig:Samplediff}
\end{figure}

A comparison between two raw combined feature segments before and after adjustment for the \texttt{add} instruction is also shown as a scatter plot in Figure~\ref{fig:Samplediff}. One can see that the CSM shifts the samples upward. For samples of the same instruction, their optimal QDA classifier boundary line (given in Equation~\eqref{formula:qda:decision}) also shifts after the adjustment. This necessitates that the classifier boundary also be updated to compensate. The comparison between non-self-enhancing classification and self-enhancing classification is achieved by measuring the recognition rate at six different time points and will be discussed in Section~\ref{sec:experiment:benchamrktesting}.

\subsection{Real-time Algorithm Implementations}
Compared with a PC platform, RASC's FPGA lacks the ability to efficiently perform complex computations and to store very large quantities of data. Thus, in a real-time implementation, we need to consider if the feature selection method, DC shift adjustment, and choice of classifier are implementable in RASC's FPGA. This section explains how each is implemented, including which need to be simplified or excluded. 

\subsubsection{Feature Selection Comparison}

As discussed in Section~\ref{sec:rt:fs}, there are 4 feature selection methods tested in the offline mode: PCA, Gini, mRMR, and filter selection. The PCA method uses all the higher dimensional (raw) features to compress them to a lower dimension, while the other three methods (Gini, mRMR, and filter selection) only use a subset of the raw features. During the processing, the PCA method needs the storage of the whole combined trace, and thus it requires too much storage and resources. Thus, it does not satisfy the requirement of real-time tasks in this paper.

\begin{figure}[t]
  \centering
\includegraphics[width=0.48\textwidth]{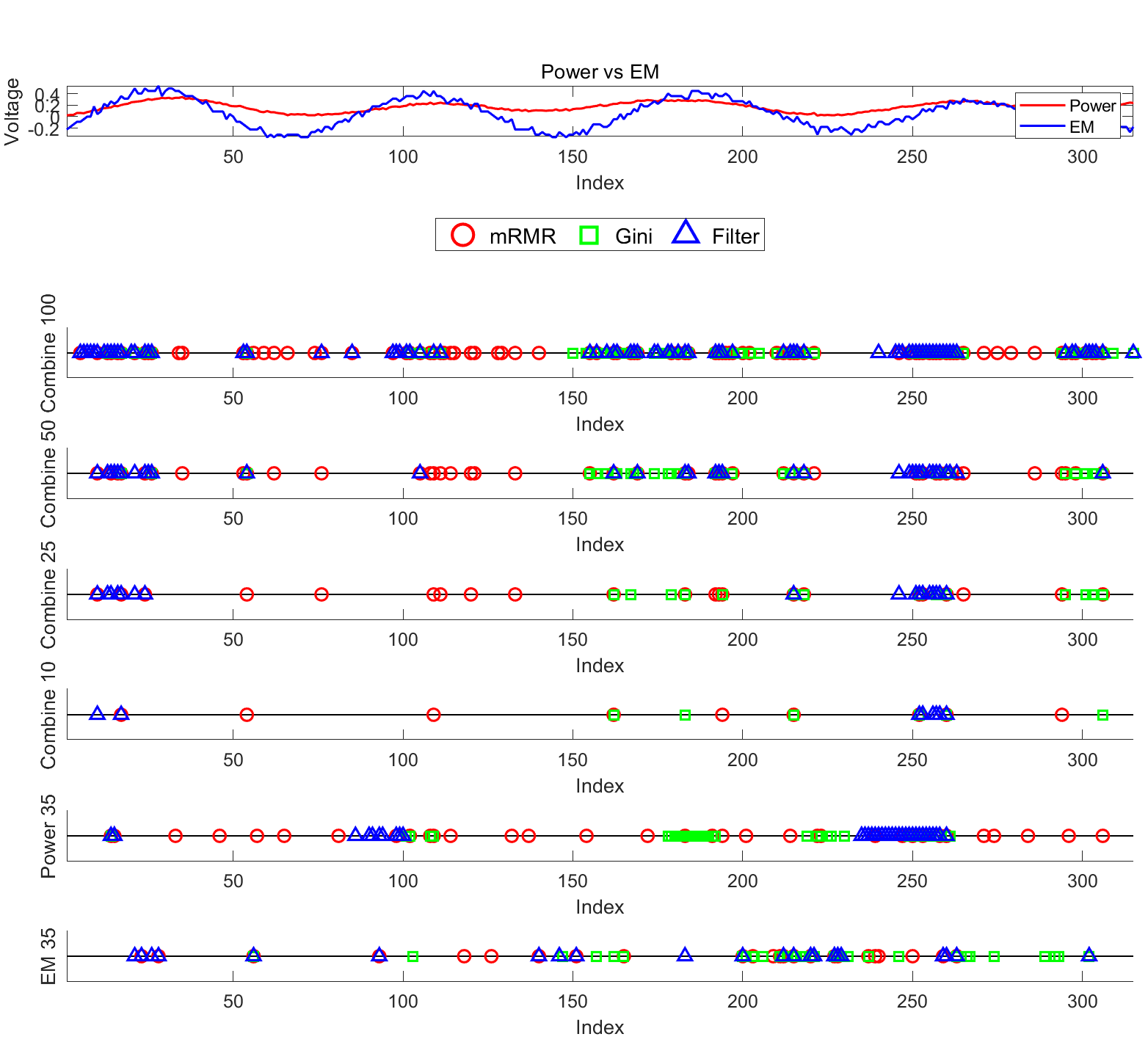} 
        \caption {Feature selection method comparison. The feature selection results within the Group 1 trace datasets using three methods: Minimum Redundancy Maximum Relevance (mRMR) (red circle), Gini index (green square), and Filter methods (blue triangle), applied to the power trace (red line) and EM trace (blue line) above.}
       \label{fig:fs_comparison}
\end{figure}

The latter three methods, however, are easier to implement, and we illustrate them as feature selection techniques in Figure~\ref{fig:fs_comparison}. This figure compares the feature selection results for the Group 1 dataset and different numbers of features. The features selected by mRMR are marked with red circles, those by the Gini index with green squares, and those by the Filter method with blue triangles. `Combine 100', `Combine 50', `Combine 25', and `Combine 10' represent the top 100, 50, 25, and 10 selected features in the combined Group 1 dual-channel traces, respectively. `Power 35' and `EM 35' indicate the top 35 selected features in the single power and EM traces. The power trace labeled 'ADD' spans two clock cycles: the first half corresponds to fetching the 'ADD' instruction, and the second half to executing it.

In the `Combine 10' plot of Figure~\ref{fig:fs_comparison}, among the ten features selected by mRMR, seven are located in the execution phase and three in the fetching phase. This distribution indicates that mRMR effectively captures significant variations across the entire power trace, ensuring a balanced and comprehensive representation. By minimizing redundancy and maximizing relevance, mRMR identifies critical points throughout the trace that are essential for effective analysis and classification in side-channel analysis. Additionally, the mRMR method selects at least one feature that overlaps with those chosen by both the Filter and Gini methods. This pattern is also observed in `Combine 100', `Combine 50', and `Combine 25', where the features selected by the mRMR method are spread across the entire trace, avoiding redundancy. On the other hand, the Gini method tends to cluster features around regions of high variation, particularly near index 180 in Figure~\ref{fig:fs_comparison}. In contrast, the Filter method selects a large number of features in close proximity, predominantly around index 250. The clustering observed in the `Combine 100', `Combine 50', `Combine 25', and `Combine 10' series suggests that the Gini and Filter methods may introduce redundancy, which can be detrimental and inefficient to disassembly tasks.

Moreover, compared to the single Power/EM channel mRMR feature selection results in `Power 35' and `EM 35', the mRMR method in the `Combine 100' plot selects features at various positions. This variation likely occurs because, even at the same time index, features from the combined channel contain more information than those from a single channel. The additional information provided by the combined channel influences the mRMR selection process, resulting in a more varied feature selection. As discussed further in Section~\ref{sec:offline}, experimental results demonstrate that the combined channel with mRMR outperforms the single channel with mRMR, as well as the combined channel with the other two feature selection methods. In summary, mRMR demonstrates superior feature selection in this context. By balancing information gain and redundancy, it leads to a more informative and less redundant set of features, making it the most effective method for this application.


\subsubsection{Self-enhancing Classifier}

 In self-enhancement, the side channels of the target device vary due to the effect of time, stability of DC power, aging of the target chip, etc. These tiny variations in the side channels might affect the accuracy of the disassembly result. The self-enhancing updates the mean and coefficient of the classifiers before the disassembly so as to increase the accuracy of the recognition rate. Equations~\eqref{formula:SA1} and \eqref{formula:SA2} require scalar multiplication, vector addition, matrix addition, and matrix multiplication which can be implemented on the FPGA.

\subsubsection{Choice of Classifier}\label{sec:meth:choiceofclassifier}


In the offline mode, the classifiers (LDA, QDA, and MLP) discussed in Section~\ref{sec:back} and many more can be implemented. However, as discussed above, real-time approaches using an FPGA will struggle with floating point calculations, large matrix multiplications and inversions, and storing huge amounts of data. 

The multilayer perceptron (MLP) is the first one that we exclude from real-time mode. Based on our offline mode experiments, at least 30 hidden layers are needed for reasonable recognition rates, which consumes too much memory for RASC. Besides, the ReLU activation layer of the MLP algorithm contains exponential calculations and is too expensive to implement on the FPGA. Compared with MLP, the QDA and LDA are more FPGA-friendly. Specifically, the coefficients in QDA and LDA can be calculated offline, and they can be translated to integers in order to avoid the floating number calculations on the FPGA. Further, the comparison of scores from different classes (required by OvA classification) could be achieved with the bubble sort method. The bubble sort method starts by comparing the score of the first and second classifier, and the higher score and their class label is saved. Then, the winner of the first and second classifiers will be compared to the third classifier, and so on, until all classes are compared. After performing preliminary experiments, we found that the QDA results were better than LDA. Therefore, we implemented QDA into RASC for real-time mode. The resources required are presented in Table~\ref{table:overhead}. The code associated with the real-time implementation will be open-sourced upon acceptance of this article.


\begin{table}[t]\setlength{\tabcolsep}{4pt}
\centering
\caption{FPGA resources used in real-time implementation.}   
\label{table:overhead}
\begin{tabular}{|c|c|c|c|c|}
\hline
\textbf{Module} & \makecell{\textbf{\# of} \\ \textbf{registers}} & \makecell{\textbf{\# of} \\ \textbf{LUTs}} & \makecell{\textbf{\# of} \\ \textbf{slices}} & \makecell{\textbf{\# of} \\ \textbf{clock cycles}} \\
\hline 
\makecell{Feature combination} & 30 & 175  & 87 & 10 \\
\hline 
\makecell{Inter classification} & 1408 & 1408 & 702 & 40 \\
\hline 
\makecell{Within-group classification} & 1408 & 1408 & 702 & 40\\
\hline 
\makecell{Bubble sort} & 12 & 258 & 145 & 20 \\
\hline 
\makecell{Self-enhancing adjustment} & 0 & 416 & 134 & 10 \\
\hline  
\end{tabular}
\end{table}

\subsubsection{Implementation of Classifiers}\label{sec:classifier_implementation}

For performing classification in real-time on RASC, the most important thing is to bypass the time-consuming calculations. The QDA classifier needs to do a vector subtraction, a matrix multiplication, a scalar addition, and a scalar comparison for each trace segment. The score of the input trace segment for
a class label $c$ is
\begin{equation}\label{formula:3}
Score_{c}(\vec{z})={-(\vec{z}-\vec{\mu}_{c})}^T \boldsymbol{\Sigma_{c}}(\vec{z}-\vec{\mu}_{c})+b_{c},
\end{equation}
where $\vec{z}$ stands for the combined input trace segment, a matrix {$\boldsymbol{\Sigma_c}$} contains coefficients that define an orthogonal vector to the hyperplane, and a scalar $b_{c}$ is the bias term of class label $c$ in the QDA binary classifier.  $b_{c}$ can be pre-calculated as 
\begin{equation}\label{formula:realtime:classifier}
b_{c}=2\left(\log(p_{c})-\frac{1}{2}\log(|\boldsymbol{\Sigma}_{c}|)\right).
\end{equation}
After calculating the outcome of all possible instructions, the target instruction is decided by the instruction number of the max output score:
\begin{equation}\label{formula:5}
instr=\mathop{\arg\max}_{c \in \{c_1,\hdots, c_n\}}\{Score_{c}(\vec{z})\}.
\end{equation} 
This comparison can be accomplished by the bubble-sort module mentioned in Section~\ref{sec:meth:choiceofclassifier}.

\subsubsection{Real-time Instruction Classification Rate}\label{sec:meth:timecost}

In real-time SCD tasks, the processing outcome of instruction $instr_i$ on RASC should be finished within the duration of instruction $instr_{i+1}$ on the target. Otherwise, the classification would not be able to keep up with the program running on  the target board, and RASC classification would only be considered as ``single-trace'' and not ``real-time''. In this paper, RASC operates at 160MS/s (or 160MHz) to sample a target running at 1MHz. In order to be classified as real-time, RASC should therefore complete its processing at a rate be faster than 1MHz, or less than 160 RASC clock cycles per instruction. The time cost of all blocks in our real-time implementation are listed in the last column of Table~\ref{table:overhead}. In RASC's real-time disassembly, the feature combination and self-enhancing adjustment cost 10 clock cycles, inter-group classification consumes 40 clock cycles, within-group classification takes 40 clock cycles, and the bubble sort uses 20 clock cycles. Thus, the  processing time per instruction is 120 clock cycles, meaning that it can classify 1.33 million instructions per second. Since $120 < 160$, RASC's disassembly can therefore be categorized as \textit{both single-trace and real-time.}

\section{Experiment setup and results}\label{sec:result}
\subsection{Experimental Setup}
\noindent \textbf{Power/EM Trace Measurements.} 
For collecting power and EM traces in offline mode, we use the MDO3102 oscilloscope\cite{MDO3102} and a commercial EM probe (LF1 Set from LANGER EMV Technik\cite{EMV}). For collecting EM and power traces in real-time modes, we use our custom-designed 38mm-by-38mm RASCv3 and a printable near-field EM antenna, and one 10 ohm shunt resistor is soldered between the GND pin of ATmega328P chip and the GND pin on Arduino UNO board for stabilizing the EM/Power leakage. Considering the difference in sampling rate between RASC (160MS/s) and oscilloscope (2.5GS/s), the core frequency of Arduino UNO is set as 1MHz and 16MHz, respectively. The power/EM traces are collected by RASC and transmitted to the PC via UART whenever offline processing is performed. In the real-time mode, RASC collects power/EM traces and processes them internally. 

\vspace{0.5ex}

\noindent \textbf{Methodology Parameters.} In our experiments, the window size of the traces ($w$) is set to 315 for  experiments, and the number of traces collected for each class to compute coefficients and mutual information ($s$) is 3,000. When testing real benchmarks, the number of selected features ($w^*$) in the real-time mode is 70, and in the  mode is 50. $w^*$ was chosen based on the recognition rates from template experiments.

\vspace{0.5ex}

\noindent \textbf{Testing Benchmarks.}
Six benchmarks~\cite{AVRcode} (see Table~\ref{table:benchmark}) are tested, and each benchmark includes one numerical function necessary for real applications such as face recognition, self-driving cars, manufacturing, etc.

 \begin{table}[t]\setlength{\tabcolsep}{4pt}
\centering
 \caption{Benchmarks used in testing.}   \label{table:benchmark}
\begin{tabular}{|c|c|c|c|}
\hline
\textbf{Benchmark} & \textbf{\# Lines} & \textbf{\# Clock cycles} & \textbf{Runtime ($\mu s$)} \\ 
\hline 
 Timeloop & 12& 1.4k & 287.5 \\ 
\hline 
 Matrix & 14& 1k& 262.5 \\
\hline 
 Decimaldivision & 10& 1.2k & 275\\ 
\hline 
 Decimal2float & 12& 1k & 262.5\\ 
\hline 
 ASCII & 16 & 2k & 325 \\
 \hline 
 ADconverter& 14 & 1.4k & 287.5 \\ 
\hline 

\end{tabular}
\end{table}

\vspace{0.5ex}

\noindent \textbf{Assumptions.} We assume the reverse engineer can access the target device and have the means to know when the testing program starts. She also knows the instruction set of the target device, 
has access to the power/EM channels, and is able to train a classifier. However, we assume that she does not know the instructions being executed by the module. This is harder than malware detection which would only require verification of instructions and control flow~\cite{bai2022real}.

\begin{figure*}[t]
  \centering
\subfloat[]{
  \includegraphics[width=0.48\textwidth]{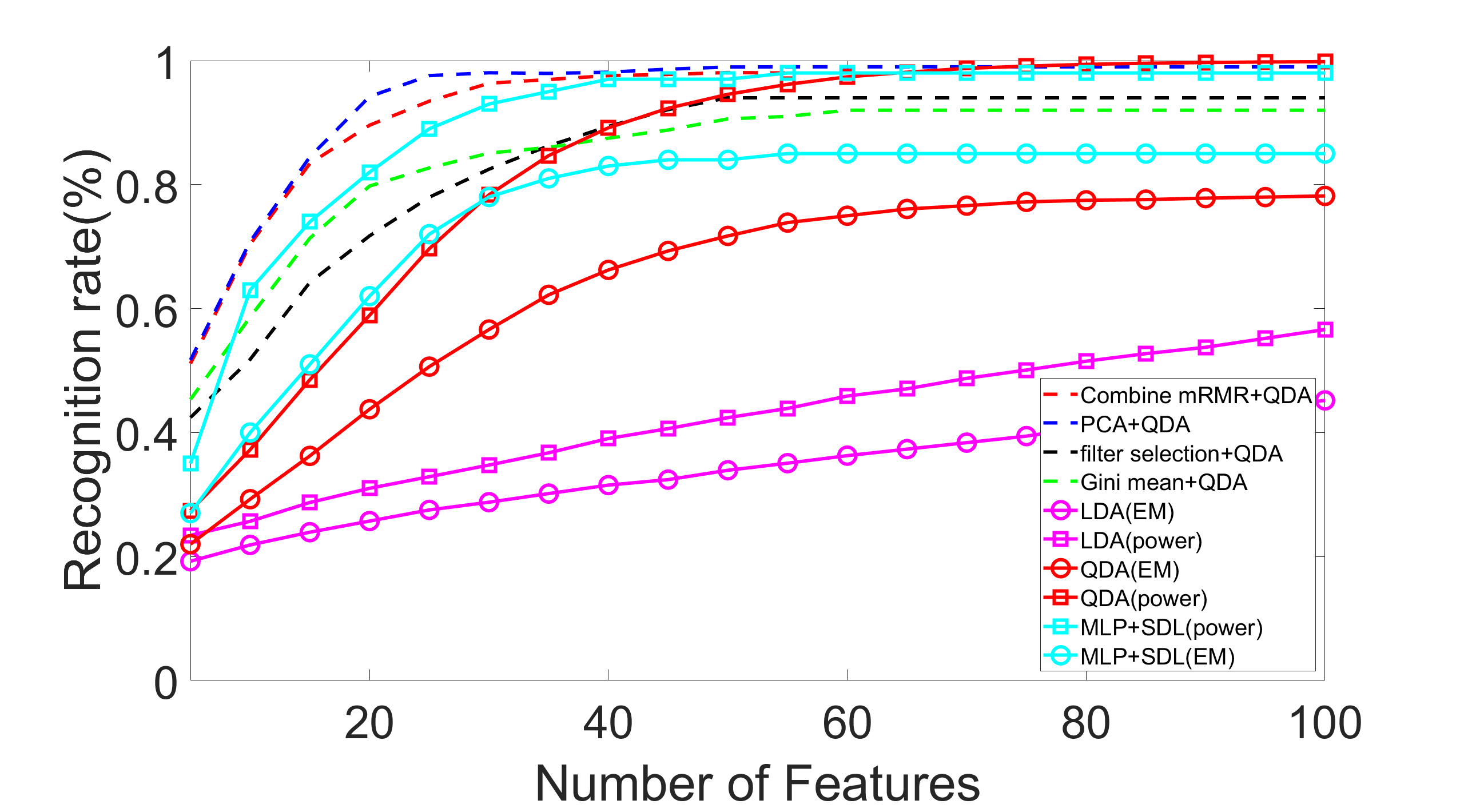}} 
\subfloat[]{
  \includegraphics[width=0.48\textwidth]{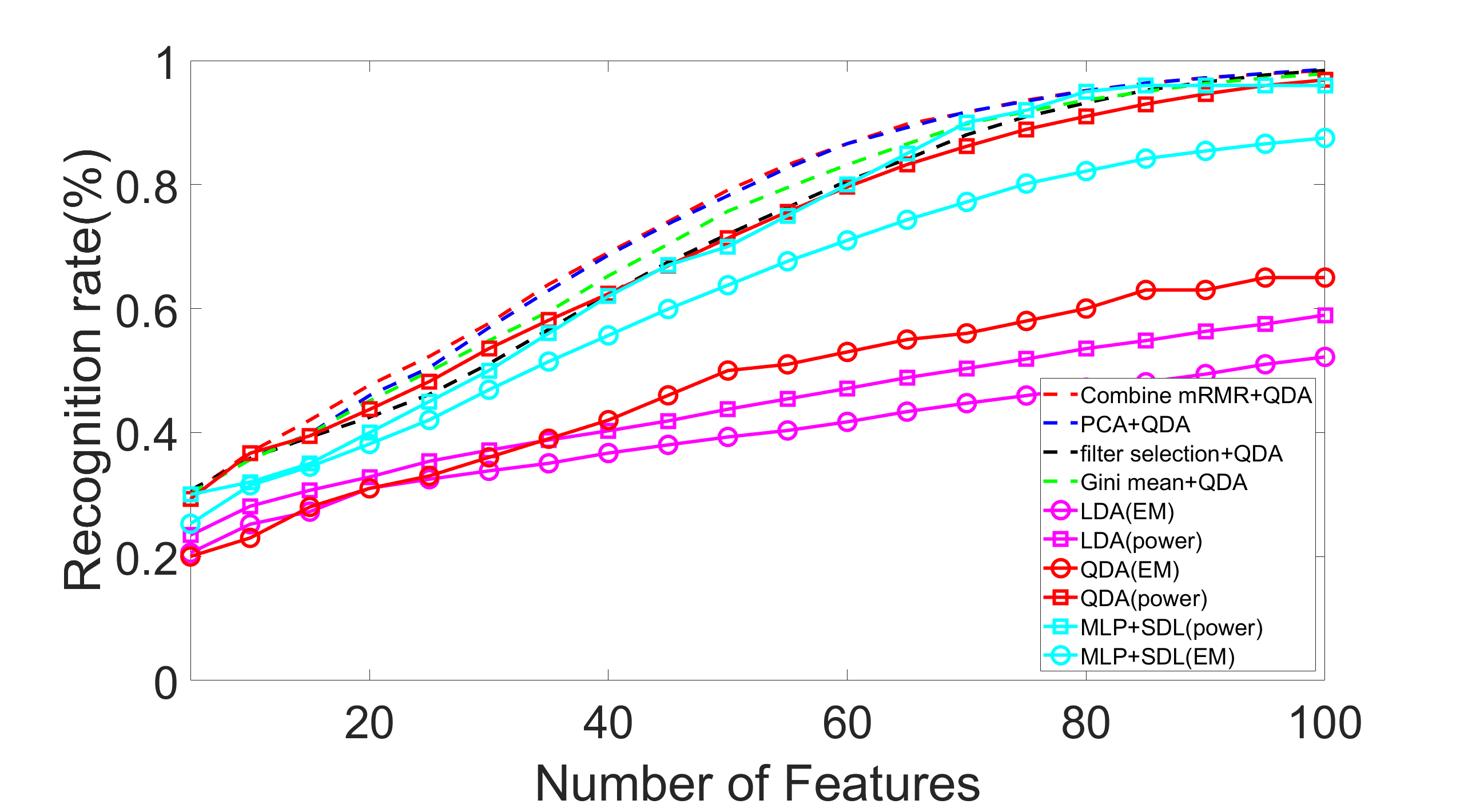}}
        \caption {Average inter-group classification QDA with mRMR/PCA/Gini/other functions from references using (a) oscilloscope and (b) RASC.}
        \label{fig:QDA_average}
\end{figure*}

\vspace{0.5ex}


\noindent \textbf{Training Templates.} The pipeline structure of the Arduino UNO is shown in Figure~\ref{fig:example}(a). One stage fetches the instruction while another pipeline executes the previous instruction. Thus, the instructions in both stages will contribute to the shape of the power and EM traces. 

The example template we use for collecting traces of target instruction is presented in Figure~\ref{fig:example}(b). It includes one \texttt{sbi} instruction for triggering the start signal, two \texttt{nop} instructions, and a random instruction, and the target instruction followed by another random instruction, two \texttt{nop}s, and one \texttt{cbi} for ending trigger signal.  The usage of the trigger signals lets RASC and oscilloscope know when to start and end collecting power traces for a given template, and the two \texttt{nop}s are used to separate the trigger signals and the target instruction. The destination register and processed data of the target instruction, as well as its preceding and subsequent instructions, are all randomly determined. 

\vspace{0.5ex}

\noindent \textbf{Pre-processing and Feature Reduction Approaches.}
The data processing methods in  mode include PCA~\cite{PCA} Gini~\cite{menze2009comparison} selection, and filter methods~\cite{yu2003feature}, and mMMR~\cite{peng2005feature,ding2005minimum}.  In real-time mode, we adopt mRMR to reduce the size of the testing traces. The algorithm for PCA could be found on Matlab's website~\cite{PCA}. 

\subsection{Classification with Example Templates}\label{sec:offline}
In the offline mode, we adopt the oscilloscope to collect traces using the example template for each instruction, randomly take 90\% of traces to build inter-group classifiers, and test the remaining 10\% of traces to get the recognition rate. The average recognition rate of all eight groups versus feature number is presented in Figure~\ref{fig:QDA_average}.

\vspace{0.5ex}

\noindent \textbf{Single vs. Combined Channel.} We start by comparing the average dual channel (EM and power) recognition rates shown in dashed lines with the single channel results (EM only or power only) shown in solid lines. Figure~\ref{fig:QDA_average} shows the double-channel result is better than the single-channel result since more information is contained in the double channel. Besides QDA and LDA, we implement the deep learning method (MLP) and SDL selection method from~\cite{fendri2022deep}. The hidden layer number is set as 30, and the activation function uses ReLU. Figure~\ref{fig:QDA_average} shows the deep learning method is better than the normal QDA/LDA classifier. Nevertheless, QDA is still easier to implement into the RASC platform and is enough accurate for instruction disassembly.

\vspace{0.5ex}

\noindent \textbf{Impact of Feature Selection.} We next compare the four different feature processing methods (PCA, Gini, mRMR, and filter) with QDA classifiers by their resulting recognition rates. For mRMR methods, we utilize the approach mentioned in Section~\ref{sec:back:mRMR} to combine EM and power features from the same time indices. For the other three processing methods (PCA, Gini, and filter), we concatenate the power and EM trace features and use them to process the features. Figure~\ref{fig:QDA_average} shows the combination of PCA and QDA is the best of the four, followed very closely by mRMR for the oscilloscope. This is plausible since mRMR methods only select the valuable part of the whole trace. On the contrary to that, the PCA method compresses the whole high-dimensional dual-channel feature vector into the low-dimensional vector, and it contains more information than other feature selection methods. However, for RASC, mRMR is better than PCA initially, and then they are neck and neck. Besides this, mRMR outperforms the other two lighter-weight feature-selection methods (Gini, filter) in the average classification results, and this entices us to adopt mRMR in real-time mode for later experiments. 

\vspace{0.5ex}

\noindent \textbf{Oscilloscope Setup vs. RASC.} In Figure~\ref{fig:QDA_average}(b) RASC collects traces for all instructions, and they are transmitted into the PC platform for classification. Compared with the average recognition rate from the oscilloscope (Figure~\ref{fig:QDA_average}(a)), RASC needs more features to achieve the same recognition rate. The oscilloscope has advantages in collecting side-channel traces due to its more sensitive ADC, broader band filter circuits, and more stable power source. Thus, we use more features in the real-time mode in later experiments.

\subsection{Benchmark Testing Results}\label{sec:experiment:benchamrktesting}


 \begin{table*}[t]\setlength{\tabcolsep}{4pt}
 \scriptsize
\centering
 \caption{Offline and real-time mode benchmark recognition rate at different time points. }  \label{table:offline}
\begin{tabular}{|c|c|c|c|c|c|c|c|c|c|c|c|c|}
 \hline
\multirow{2}{*}{\diagbox{\textbf{Benchmark}}{\textbf{Time point}}} & \multicolumn{2}{c|}{1st} & \multicolumn{2}{c|}{2nd} & \multicolumn{2}{c|}{3rd} & 
\multicolumn{2}{c|}{4th} & \multicolumn{2}{c|}{5th} & 
\multicolumn{2}{c|}{6th}  
\\ \cline{2-13} & Offline &Real-time & Offline &Real-time & Offline &Real-time & Offline &Real-time & Offline &Real-time & Offline &Real-time 
\\ \hline
\textbf{Timeloop} & 80\% &72\% & 85\%& 76\% &85\% &76\% & 93\% &78\% &93\% &78\%  &93\% &78\%
\\ \hline
\textbf{Matrix} &83\% &71\%  & 90\% &78\% &89\%&75\% & 90\% & 80\% &90\% & 80\% & 90\% &80\% 
\\ \hline
\textbf{Decimaldivision} &74\% &72\%   &81\% &76\% &80\%  &74\% 
&89\%  &80\% &88\%  &80\% &93\% &80\%
\\ \hline
\textbf{Decimal2float} &78\% &76\% &85\%  &80\% &83\% &78\%  & 87\% &81\%& 87\% &80\%  & 91\% & 82\%
\\ \hline
\textbf{ASCII} & 75\% &65\%  &80\% &74\% &79\% &75\%  &86\%  &80\%  & 85\% & 77\% & 89\% & 81\%
\\ \hline
\textbf{ADconverter} & 78\% &70\% &85\% &80\%  &85\% &78\%  & 85\% &80\%  & 85\% &80\% & 87\% &81\%
\\ \hline \hline
\textbf{Avg.} & 78\% &71\% &84\% &77\%  &83\% &76\%  & 88\% &80\%  & 88\% &79\% & 90\% &80\%

\\ \hline 

\end{tabular}
\end{table*}

\begin{figure}[t]
  \centering{
  \includegraphics[width=0.48\textwidth]{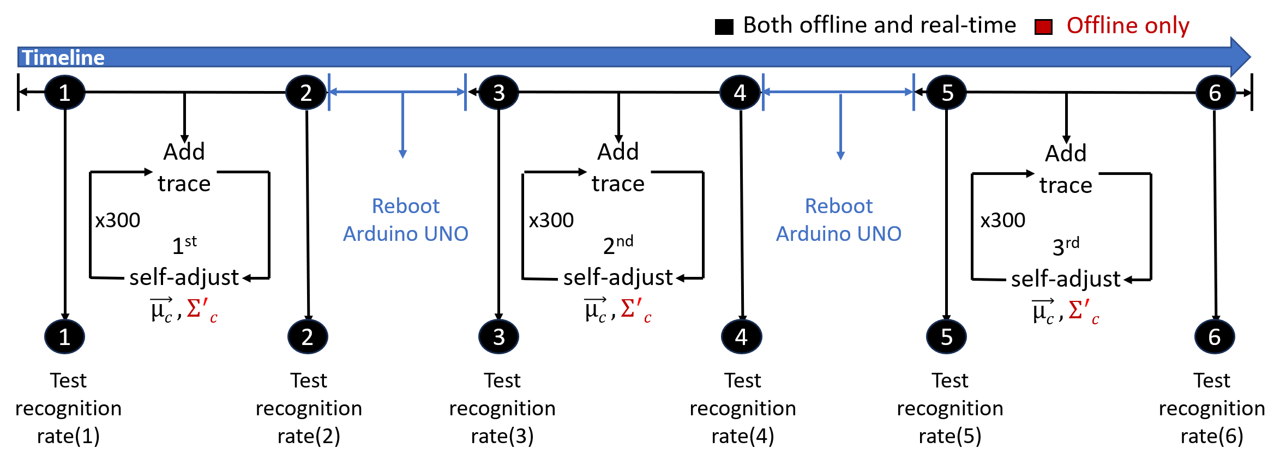}}
        \caption{Six time points used in the benchmark testing. Offline mode updates covariance matrix throughout the process while real-time mode does not. Both modes employ the self-enhancing classifier at times 2, 4, and 6 to update classifiers using the next 300 traces.}
        \label{fig:timeline}
\end{figure}

In this section, we perform disassembly on the classifiers trained on templates using real benchmark codes and traces. To capture the improvements obtained from our covariate shift adjustment schemes, we compare the recognition rates at six-time points.
The details of the timeline are presented in Figure~\ref{fig:timeline}. Between the first and second time points, the recognition rate is obtained without the self-adjustment. After the second time point, there are 400 trace segments that are used for the covariate shift minimization algorithm and self-adjustment of the QDA classifier coefficients. The recognition rate after the second time point is tested after the adjustment of the classifier. Later, to imitate the DC shift caused by different times and system reboot, we restart the Arduino UNO board and perform recognition at the third time point (before new adjustments can take effect) and then again at the fourth time point (after the adjustments can take effect). The restart is repeated, and recognition rates are collected again at the fifth and sixth time points under the same conditions. 


\vspace{0.5ex}

\noindent \textbf{Offline Mode.} We first examine the offline mode. As shown in Table~\ref{table:offline}, the average disassembly recognition rate rises from 78\% in the first time point on average to 90\% at the sixth time point. At every reboot, the disassembly results are affected. However, the self-adjustment improves the results over time. The disassembly recognition rate increases from the first, third, and fifth-time points after adjustments. Besides, the overall recognition rates and improvements differ by the instructions used in each benchmark. For instance, Timeloop's recognition rate reaches 93\% after the fourth time point, while Matrix's recognition rate reaches around  90\% after just the second time point. The recognition rates of several other benchmarks are still improving at the sixth point and might improve even more with more traces.


\vspace{0.5ex}
\noindent \textbf{Real-time Mode.} In the real-time mode, we implement the self-adjusting QDA classifiers into RASC. As mentioned in Section~\ref{sec:meth: real}, we only update the mean value of each instruction of the classifiers when using self-adjustment in this mode. Despite this, the recognition rate across benchmarks is still 80\% on average after the six-time point. Further, the self-adjustment alone still shows improvements in the recognition rate after the adjustment (1st vs. 2nd, 3rd vs. 4th, 5th vs. 6th). In comparison to the results obtained in offline mode, the recognition rate in real-time mode is lower, even when utilizing a greater number of features (50 vs. 70). One possible reason is the approximation in the classification decreases the recognition accuracy. Besides, the commercial oscilloscope has higher sensitivity and a better current filter circuit than RASC, which could let the collected EM and power traces be more stable than RASC. Nevertheless, RASC combined with the proposed algorithms still shows impressive potential in real-time disassembly and (potentially) malware detection.

\subsection{Supplementary Results}

This article has a supplementary document with recognition rates for all inter-group and intra-group classifications in confusion matrices.

\section{Discussion} \label{sec:complex_targets}

Given the cost and complexity of designing a high-speed version of RASC and the fact that we were not sure if real-time disassembly would even be possible before our attempts to do so, the sampling speed of RASCv3 was intentionally designed to be low (160MS/s), and the target chosen was the Arduino UNO for proof-of-concept at low cost.
However, the proposed approach is quite generic and should be applicable to more complex targets assuming better measurement setup (e.g., sampling rate, SNR, etc.) and/or processing resources (e.g., DSP). 
The same overall flow could be used regardless of the target. 
Furthermore, it is intuitive that the proposed dual-channel approach, which combines the best features from power and EM channels, would outperform single-channel approaches for other targets. 
In future work, we plan to test the framework on more advanced MCUs with 3 or 5 pipeline stages or multicore architectures. 
In this section, we summarize the challenges associated with more complex targets, as identified in recent work~\cite{maillard2023side,iyer2024hierarchical,krishnankutty2020instruction,yilmaz2022marcnnet} and our own experiments. We also outline potential improvements to enhance RASC and our methods for real-time SCD of these more complex targets.   
\subsection{Concurrency and Parallel Execution}
Pipelining and multi-issuing\cite{maillard2023side} are inevitable challenges when disassembling complex targets. Pipelining decomposes the execution of instructions into an assembly line, and multi-issuing allows multiple instructions to execute in parallel. Similarly, multi-core processors could have multiple pipelines operating in parallel. All three of these will cause the overlap of side-channel leakages of multiple instructions being executed at the same time, which will lower the correlation between side channels and instruction labels. The accuracy of any classifier is likely to decrease as a result. 

There are a few ways to combat this. First, more capable classification algorithms that use floating point arithmetic, matrix multiplications and inversions, frequency and wavelet transformations, convolutions, and activation functions may be used in place of QDA. This shall require upgrading RASC's processing capabilities to include digital signal processing (DSP) that works within or alongside the FPGA. Next, the SNR of RASC's EM measurement setup could be improved with new materials, an upgrade to an array of smaller antennas, etc. Our experiments already showed the benefits reaped from combining EM and power features to increase the correlation with instructions. Measurements from additional EM antennas represent additional channels from across the target and could be correlated with different cores and instructions being run in parallel. Those channels could be combined using mutual information just like in this article to select the subset that offers the best classification rate per instruction. In addition, mRMR finds the best places to sample from within the combined traces of each instruction.

\subsection{Trace Slicing}
Most of the Arduino UNO instructions only takes one clock cycle and a few of them take two. Some MCUs, such as AT89S51\cite{iyer2024hierarchical}, have instructions with three kinds of clock durations, e.g., 1, 2, and 4 clock cycles. For each case, the number of bytes fetched from program memory varies. This increases the complexity of slicing a predefined window of traces into pieces since there exists more possible instruction combinations. For example, a 6-clock cycle trace has 10 possibilities if we only take 1 or 2 clock cycles into consideration. However, if we also take 4-clock-cycle instructions into account, there exists 15 possibilities. One needs to add a phase to analyze the most possible instruction combinations from all potential possible instructions as mentioned in \cite{krishnankutty2020instruction}. 

\begin{figure}[t]
    \centering    \includegraphics[width=0.48\textwidth]{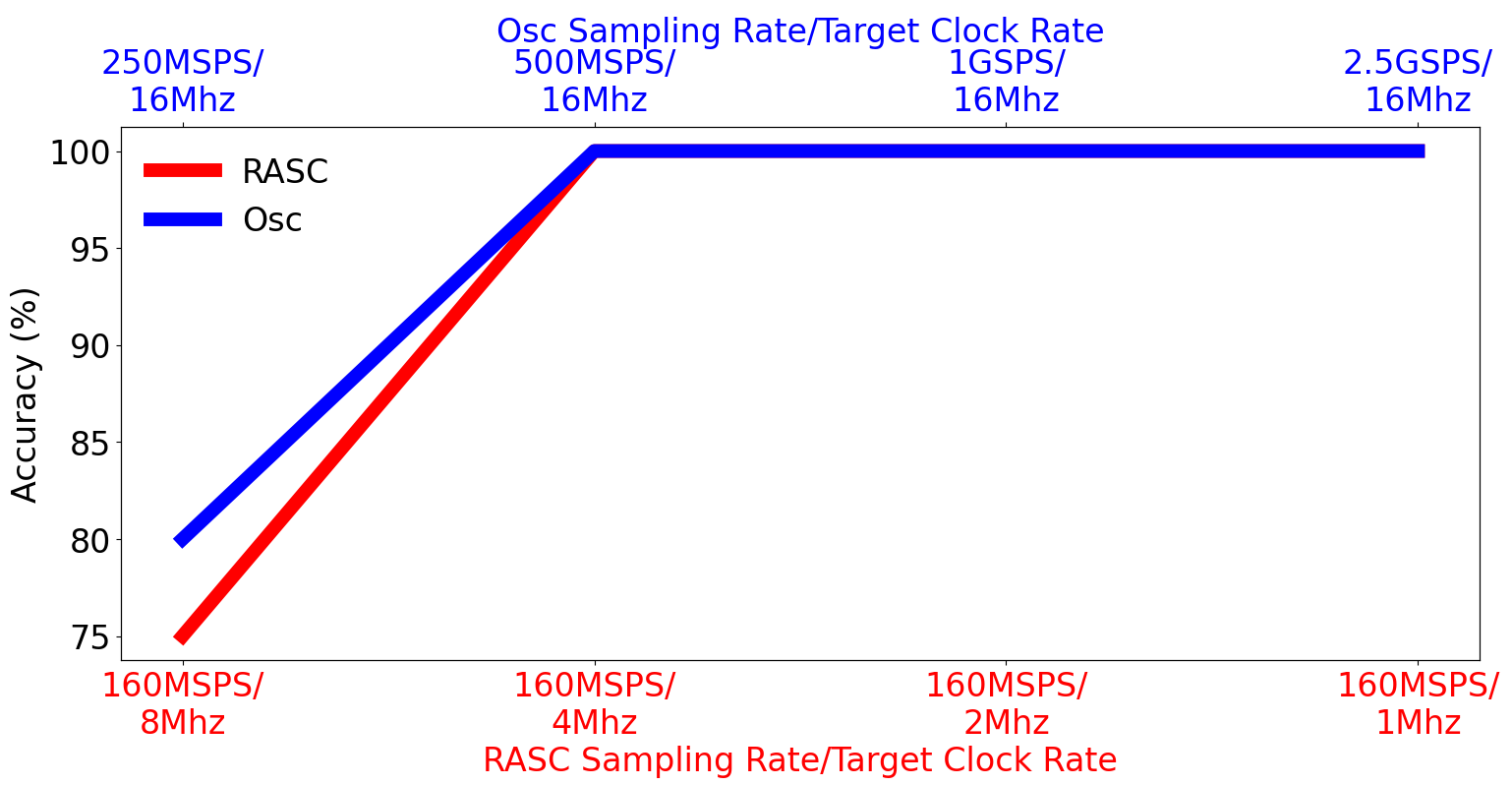}
    \caption{Max recognition rate versus sampling points per clock cycle.}
    \label{fig:samplediff}
\end{figure}

\subsection{Target Clock Frequency}
The clock speed of MCUs used in OT systems is faster than an Arduino UNO, and this brings challenges to next version RASC and other real-time disassembly platforms. In this article, RASCv3 samples 160 points per clock cycle (160MS/s for SCD of a 1MHz core). If the core frequency of the target increases to 100MHz, like STM B-L475E, the sampling speed of RASC should be 1.6GS/s to maintain the same number of features. Faster ADCs are available but will increase the size and cost of RASC. Thus, in the next version, we aim to find the balance between the sampling speed of RASC and size/cost for a given application case. For example, when using RASC to monitor an IoT system, sampling at such high rates might not be cost effective; however, for OT systems that are large, more expensive, and would have more dire consequences if attacked, the added cost could be more than reasonable.

In Figure~\ref{fig:samplediff}, we show the result of a set of tests on the maximum recognition rate versus sampling rate. In the top x-axis, the offline setup's sampling rate is increased while the target's (Arduino UNO's) clock rate is fixed. In the bottom x-axis, RASC's sampling rate is held fixed while the Arduino's clock frequency is inceased. For the Arduino UNO, these results suggest that when using an oscilloscope, capturing 30 to 35 features per clock cycle is sufficient for successful disassembly. RASC, on the other hand, requires slightly more features per clock cycle -- 35 to 40 -- to reach the same level of accuracy. Looking ahead to future work, particularly when targeting more complex MCUs like the STM32, the challenges of signal overlapping must also be considered. However, it's important to note that instructions for such MCUs typically take more clock cycles compared to the Arduino UNO (most take 1 clock cycle). This means there will be more features available per instruction, potentially simplifying SCD despite the increased complexity. Thus, for sampling 100MHz target, 4GS/s may be good enough for RASCv4. Based on the corresponding ADC listed on Digikey, the total cost could still be lower than \$4,000, which is a fair amount compared to offline setups, such as MDO3102 (5GS/s at \$15,000).

\subsection{Dynamic Voltage and Frequency Scaling }

Dynamic Voltage and Frequency Scaling (DVFS) is a power management technique where the voltage and frequency of a processor are dynamically adjusted based on workload. It may be utilized in SoCs and some MCUs. DVFS introduces significant variations in power consumption and EM, which can distort the signal patterns exploited by SCD. Additionally, changes in operating frequency affect the timing and execution speed of instructions, complicating the alignment and interpretation of data. This can lead to inaccuracies in classifiers that assume stable operational conditions. Moreover, DVFS can lead to non-linear and unpredictable power profiles, further complicating the establishment of baselines.

To improve the SNR of side-channel signals, sophisticated signal processing algorithms may be added to filter out noise and normalize data for variations introduced by DVFS. For synchronization, algorithms are needed to ensure the measurement process is aligned with the target device's operating state. Furthermore, developing predictive models to anticipate how DVFS settings change in response to different workloads will help adjust SCD classifiers accordingly.

\section{Related work}\label{sec:rw}
The related work section will first introduce the EM-based and power-based disassembly papers. Then dual-channel side-channel attack papers are also discussed. All of them are compared to the proposed approach in Table~\ref{table:relatedwork}. 

 \begin{table*}[t]\setlength{\tabcolsep}{4pt}
 \scriptsize
\centering
 \caption{Comparison of state of the art with the proposed method. }\label{table:relatedwork}
\begin{tabular}{|c|c|c|c|c|c|c|c|c|c|}
\hline
\diagbox[]{\textbf{Criteria}}{\textbf{Reference}}
 & \textbf{\cite{strobel2015scandalee}} & \textbf{\cite{cristiani2020bit}} & \textbf{\cite{vaidyan2020instruction}}  & \textbf{\cite{msgna2014precise}} & \textbf{\cite{park2018power}} & \textbf{\cite{fendri2022deep}} & \textbf{\cite{krishnankutty2020instruction}} & \textbf{\cite{eisenbarth2010building}}   & \textbf{Proposed} \\ 
 \hline
\textbf{Modality} &EM  & EM & EM &   Power & Power &Power  & Power  &Power  & Power+EM  \\ 
\hline
\textbf{\# of Instructions} &35  &35  & 39 & 39   & 86 & 34 & 152  &33  & 86  \\ 
\hline
\textbf{Classifiers} & \makecell{LDA} & \makecell{QDA} &\makecell{CNN}  & \makecell{PCA+KNN} & \makecell{QDA, LDA, SVM}& \makecell{MLP+SDL} & \makecell{SVM} & \makecell{LDA} & \makecell{QDA+mRMR} \\   
\hline
\makecell{\textbf{Platform}} & \makecell{PIC16F687} & \makecell{ PIC16}  &\makecell{atmega328P }   & \makecell{atmega163} & \makecell{atmega328P} &\makecell{RISC-V processor}  & \makecell{MSP430}  &\makecell{PIC16F687}  & \makecell{atmega328P} \\    
 \hline
\textbf{Mode} & Offline  & Offline   &Offline & Offline  &Offline  &Offline  &Offline &Offline  & \makecell{Real-time} \\
 \hline
\makecell{\textbf{Covariate Shift Adjustment}} & No  & No   &No & No  &No  &No  &No &No  & Yes \\
 \hline
\makecell{\textbf{Real Benchmarks}} &Yes   & No   &No & No  &No  &No  &No &No  & Yes \\
 \hline
 
\hline
\end{tabular}
  \label{table:comparison}
\end{table*}

\vspace{0.5ex}

\noindent \textbf{EM-based Disassembly.} In~\cite{strobel2015scandalee}, Strobel et al. present a side-channel-based disassembler using localized EM emanations. They combine EM traces from multiple measurement positions and train an LDA-based classifier for PIC chip instructions. To improve the efficiency of the classification, Strobel concatenates the traces from several positions belonging to the same class, expecting the dimensionality reduction algorithm to extract the most relevant features. The trained LDA classifier could reach a 96.24\% recognition rate on the test set and 87.69\% on real code. In~\cite{cristiani2020bit}, Cristiani et al. build a side-channel disassembler to recover the bit encoding of an instruction inside a PIC16F chip using local EM leakage. To improve recognition rate and efficiency, they propose a greedy algorithm that determines the best subset of positions to predict a target bit. The proposed disassembler could achieve 99.41\% on a bit level and 95\% on full 14-bit instructions. In~\cite{vaidyan2020instruction}, Vaidyan et al. developed an EM spectral domain framework and successfully reduced dimension and feature selection using PCA. Different algorithms (QDA, SVM, and CNN) are considered in their classification. The experiment results show over 99\% accuracy in identifying AVR instruction from the ATmega328P chip. In \cite{yilamz2019instruction}, Yilam et al. adopt the upsampling method to track instructions inside an advanced microprocessor whose core frequency is too high to sample. In the training phase, the authors generate the pseudo-code, which is in the sequence of empty-loop, target instructions, and empty-loop, and collect EM signatures under low sampling speed from the target device. Then, the low-sampling EM traces are adjusted to high-sampling EM traces with a modulo operation. The experimental results show a high correlation between the EM traces after the modulo operation and the EM traces after a single execution. 

Compared with these papers, the proposed methodology includes a special self-adjustment mechanism for dealing with covariate shifts and is suitable for real-time implementation. Further, it combines EM and power.

\vspace{0.5ex}

\noindent \textbf{Power Disassembly.}
In \cite{msgna2014precise}, Msgna et al. adopt the $k$-nearest neighbor method and collect traces from the ATMega163 chip for all AVR instructions. The Euclidean distance function is used to calculate the distance between the testing trace and the training trace. Then, features are selected by checking the sum of the difference of means, and the SNR is improved by analyzing the probability of noise in the power traces. The successful recognition rate could be up to 100\% with k$\geq$13 in the kNN model. Park et.al~\cite{park2018power} propose a hierarchical classifier by dividing all AVR instructions into eight groups and generating inter-class and within-class classifiers. The PCA method and KL divergence method are chosen to reduce the dimension of the dataset. In the end, the successful recognition rate of the classifier could be up to 100\%. In \cite{fendri2022deep}, Fendri et al. developed a deep learning (DL) framework to disassemble instructions inside a CPU. They design a model to calculate the static and dynamic power consumption at each clock cycle. Then, the DL-based deep learning framework is trained, and its features are minimized using dictionary learning with sparse coding. The proposed framework could achieve 96\% accuracy for RISC-V instructions. In \cite{krishnankutty2020instruction}, Krishnankutty et al. design both fine-grained and coarse-grained SVM-based classifiers for identifying several continuous instructions. The author first divides all combinations of instructions into 18 classes. The proposed SVM classifier is generated with power traces from multiple ports on the target device. The trained coarse-grained and fine-grained classifier is verified by the cross-validation method, and the successful detection rate for unknown combinations of instructions is close to 100\%. In the end, the author compares the coarse-grained and fine-grained classifiers by analyzing the mutual information between features and classes. The experiment result shows the fine-grained classifiers have larger mutual information and entropy than the coarse-grained method. 

Compared with these papers, the proposed methodology corrects covariate shifts and considers real-time applications. Besides, we not only examine the recognition rate of traces collected under an example template but also check the recognition rate of code in real benchmarks. 

\vspace{0.5ex}

\noindent \textbf{Dual Channel Side-Channel Attacks.}
In \cite{standaert2008using}, Standaert et al. present fair information and security metrics to evaluate the EM and power channel. The result demonstrates the EM channel has significantly higher information leakage. Then, they concatenate the power and EM channels. The experimental results show that the conditional entropy of the concatenated trace could be diminished, which improves key recovery from cryptographic devices. In \cite{souissi2012towards}, Souissi et al. first combine Pearson and Spearman correlation coefficient distinguishers for side-channel attacks. Afterward, the CPA efficiency is shown to improve by 50\%. Then, they combine EM and power traces with the square over standard deviation method, which increases the SNR of the traces. With this combined trace, CPA efficiency increases by 45\%. 
Unlike the fixed combination of \cite{souissi2012towards} and concatenate of \cite{standaert2008using}, the proposed methodology has higher granularity in selecting and combining features. 

\section{Conclusion and Future Work}\label{sec:conclusion}
In this article, we proposed the first dual side-channel instruction disassembler. The proposed approach combined and selected features based on mutual information and included two self-adjustment schemes to resolve covariate shifts. After simplifying the proposed algorithms, they were included in a small RASC platform and tested for real-time applications. With the classification result from RASC, the recognition rate of the self-adjusted classifier could achieve 99\% in example benchmark testing, 90\% in offline mode real benchmark testing, and 81\% in real-time mode benchmark testing. In future work, we plan to upgrade RASC to a more capable version, which may utilize DSP modules and/or chips. The planned upgrade could not only improve the recognition rate but also help implement more complex classifiers. Besides, we also plan to analyze the probability of individuals and sequences of instructions, like the Hidden Markov Model in~\cite{eisenbarth2010building}. The groups selected can also be optimized to improve hierarchical classification.
 
\section*{Acknowledgments}
This work has been supported in part by the US Army Research Office (ARO) under award \# W911NF-19-1-0102.


\bibliographystyle{IEEEtran}
\bibliography{reference/MainTDSC}

\begin{IEEEbiography}[{\includegraphics[width=1in,height=1.25in,clip,keepaspectratio]{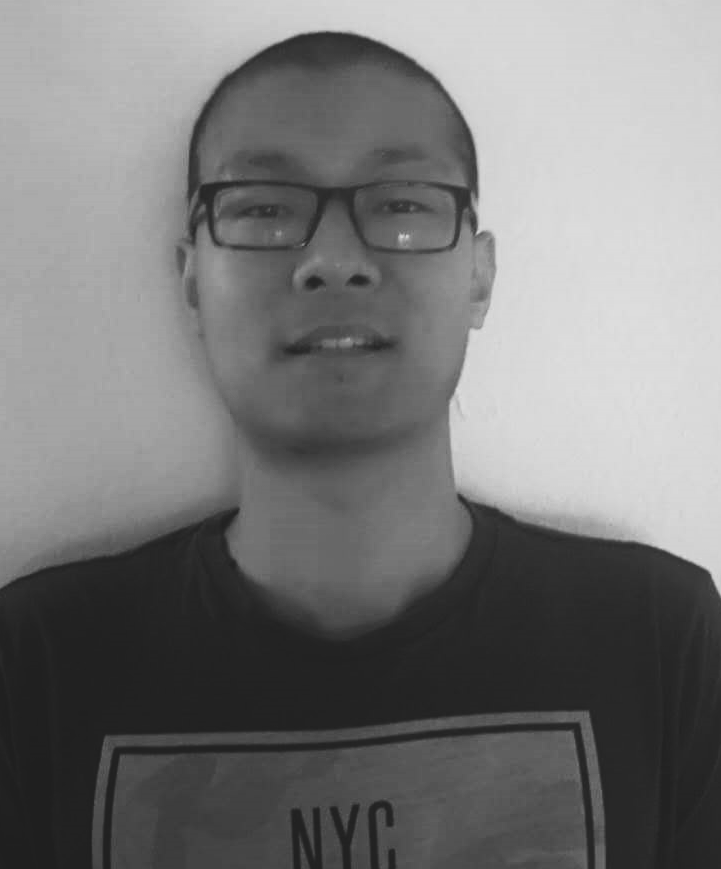}}]{Yunkai Bai} received the B.S. degree in electrical engineering from the HeiLongJiang University, China in 2017, and received the M.S. degree in electrical computer engineering from University of Florida, Gainesville, FL. Currently He is a Ph.D. candidate in the Department of Electrical and Computer Engineering, University of Florida, Gainesville, FL. His research interests focus on the domain of side-channel analysis, PUF design, and printable circuit fabrication. 
\end{IEEEbiography}

\vskip -2\baselineskip plus -1fil

\begin{IEEEbiography}[{\includegraphics[width=1in,height=1.25in,clip,keepaspectratio]{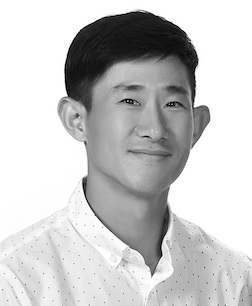}}]{Jungmin Park}  
received his B.S. degree and M.S. in electrical engineering from Kyunghee University, Korea, in 2007, and the M.S. in computer engineering from Kyunghee University, Korea, in 2009, and Ph.D. in computer engineering from Iowa State University, Ames, IA, in 2016. He is currently a senior security penetration test engineer at Lucid Motors, Newark, CA, USA. His research interests include side-channel disassembly, side-channel attacks (SCAs), SCA-resistant hardware design, and fault injection attacks.  
\end{IEEEbiography} 

\vskip -2\baselineskip plus -1fil

\begin{IEEEbiography}[{\includegraphics[width=1in,height=1.25in,clip,keepaspectratio]{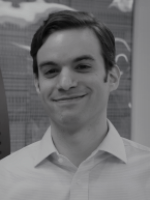}}]{Domenic Forte} received
the B.S. degree from the Manhattan College, Riverdale, NY, USA, in 2006, and the M.S. and Ph.D. degrees from the University of Maryland at College Park, College Park, MD, USA, in 2010 and 2013, respectively, all in electrical engineering. He is currently a Professor with the Electrical and Computer Engineering Department, University of Florida, Gainesville, FL, USA. His research interests include the domain of hardware security, including the investigation of hardware security primitives, hardware Trojan detection and prevention, electronics supply chain security, and anti-reverse engineering. He was a recipient of the Presidential Early Career Award for Scientists and Engineers (PECASE), the NSF Faculty Early Career Development Program (CAREER) Award, and the Army Research Office (ARO) Young Investigator Award. 
\end{IEEEbiography}

\vfill
\newcommand{\beginsupplement}{
       \setcounter{table}{0}
        \renewcommand{\thetable}{S\arabic{table}}
        \setcounter{figure}{0}
        \renewcommand{\thefigure}{S\arabic{figure}}
     }

\clearpage
\onecolumn

\section{Supplementary Material}
In this supplemental material, we present all the inter-group and within-group recognition rates of all instructions in the offline and real-time modes. Table~\ref{table:group} provides the division of Arduino UNO instructions into 8 groups. The classifiers are trained with the training traces collected using oscilloscope in the offline mode. In real-time mode, all data collection and processing are performed using RASC. For testing traces, we randomly permute all instructions from one same group to generate within-group benchmarks, and disassemble these instructions with combined traces. For inter-group classification, we adopt the example template to randomly select instructions from 8 groups to generate inter-group code segment. Then, we collect the dual-channel traces and disassembly the group number of them. In all the tables, the offline mode recognition rate of instructions from 8 groups are listed out of the brackets, and the real-time mode recognition rate of instructions are listed within brackets. In Table~\ref{tab:group_rec}, we provide the recognition rates for inter-group classification in a confusion matrix. The remaining tables provide the recognition rates in confusion matrices for every group's within-group classification.

\setcounter{table}{0}
\renewcommand{\thetable}{SM\arabic{table}}







 \begin{table}[ht]
 \scriptsize
 \centering
 \caption{Offline  (real-time) Inter-group disassembly result} \label{tab:group_rec}
\begin{tabular}{|c|c|c|c|c|c|c|c|c|}
 \hline
{} & {Group 1} & {Group 2} & {Group 3} & {Group 4} & {Group 5} & {Group 6} & {Group 7} & {Group 8}  \\
\hline
Group 1 & 0.94 (0.92) & 0 (0.01) &0 (0.01) & 0 (0) & 0.01 (0.02) & 0.01 (0.01) & 0.02 (0.03) & 0.02 (0.05)  \\
\hline
Group 2 & 0 (0) & 0.98 (0.95) & 0 (0) & 0 (0) & 0 (0.04) & 0 (0) & 0 (0) & 0 (0)  \\
\hline
Group 3 & 0 (0.01) & 0 (0) & 0.95 (0.9) & 0 (0) & 0 (0.01) & 0.01 (0.02) & 0.02 (0.02) & 0.02 (0.04) \\
\hline
Group 4 & 0 (0) & 0 (0) & 0 (0) & 0.95 (0.91) & 0 (0) & 0.01 (0.01) & 0.02 (0.04) & 0.02 (0.03)\\
\hline
Group 5 & 0 (0) &0 (0) & 0 (0) & 0 (0) & 1 (0.99) & 0 (0.01) & 0 (0) & 0 (0)  \\
\hline
Group 6 & 0 (0) & 0 (0) & 0 (0) & 0 (0) & 0 (0) & 0.99 (0.98) & 0.01 (0.01) & 0 (0.01) \\
\hline
Group 7 & 0 (0) & 0 (0) & 0 (0) & 0 (0) & 0 (0) & 0 (0) & 0.98 (0.96) & 0.02 (0.03)\\
\hline
Group 8 & 0 (0) & 0 (0) & 0 (0) & 0 (0) & 0 (0) & 0 (0.01) & 0.01 (0.01) & 0.99 (0.98) \\
\hline
\end{tabular}
\end{table}

\newpage

\begin{table}[h]
 \centering
 \rotatebox{90}{
 \begin{minipage}{1.15\textwidth}
 \scriptsize
 \caption{Offline  (real-time) within-group 1 instruction disassembly result} 
\begin{tabular}{|c|c|c|c|c|c|c|c|c|c|c|c|c|c|}
 \hline
{} & {ADC} & {ADD} & {AND} & {CP} & {CPC} & {CPSE} & {EOR} & {MOV} & {MOVW} & {OR} & {SBC} & {SUB} \\
\hline
ADC & 0.9 (0.82) & 0.01 (0.01) & 0.01 (0.02) & 0.02 (0.02) & 0 (0.01) & 0.01 (0.03) & 0 (0.02) & 0.01 (0.02) & 0 (0) & 0.01 (0.01) & 0.01 (0.01) & 0.01 (0.03) \\
\hline
ADD & 0.02 (0.03) & 0.89 (0.76) & 0.02 (0.02) & 0.02 (0.03) & 0 (0.01) & 0.01 (0.03) & 0.01 (0.02) & 0.01 (0.02) & 0 (0.01) & 0.01 (0.01) & 0.01 (0.03) & 0.01 (0.02) \\
\hline
AND & 0.01 (0.02) & 0.04 (0.03) & 0.88 (0.82) & 0.02 (0.03) & 0.01 (0.01) & 0.01 (0.01) & 0.02 (0.02) & 0 (0) & 0 (0) & 0.01 (0.01) & 0 (0.03) & 0.01 (0.01) \\
\hline
CP & 0.01 (0.02) & 0.01 (0.02) & 0.02 (0.01) & 0.87 (0.79) & 0.02 (0.02) & 0.02 (0.02) & 0.02 (0.02) & 0.01 (0.02) & 0 (0) & 0.01 (0.02) & 0.01 (0.03) & 0.01 (0.02) \\
\hline
CPC & 0.01 (0.01) & 0.01 (0.02) & 0.01 (0.02) & 0.01 (0.01) & 0.88 (0.79) & 0.01 (0.03) & 0.01 (0.03) & 0.01 (0.01) & 0.01 (0.02) & 0.02 (0.02) & 0.02 (0.01) & 0.02 (0.03) \\
\hline
CPSE & 0 (0.04) & 0.02 (0.01) & 0.01 (0) & 0.01 (0.02) & 0.03 (0.02) & 0.86 (0.79) & 0.01 (0.02) & 0.01 (0.03) & 0.01 (0.01) & 0.01 (0.02) & 0.01 (0.02) & 0.01 (0.01) \\
\hline
EOR & 0.01 (0.02) & 0.01 (0.03) & 0.02 (0.02) & 0.02 (0.02) & 0.02 (0.04) & 0.01 (0.02) & 0.87 (0.77) & 0.01 (0.02) & 0 (0.02) & 0.01 (0.01) & 0.01 (0.02) & 0.01 (0.01) \\
\hline
MOV & 0.02 (0.03) & 0.01 (0.02) & 0.01 (0.02) & 0.01 (0.03) & 0 (0.01) & 0.01 (0.01) & 0 (0.01) & 0.89 (0.79) & 0 (0.03) & 0.02 (0.03) & 0.01 (0.01) & 0.01 (0.02) \\
\hline
MOVW & 0 (0) & 0.01 (0.02) & 0.01 (0.01) & 0.02 (0.02) & 0.01 (0.02) & 0.01 (0.01) & 0 (0) & 0.02 (0.03) & 0.86 (0.76) & 0.03 (0.08) & 0.03 (0.03) & 0 (0.01) \\
\hline
OR & 0.01 (0.03) & 0 (0.01) & 0 (0.01) & 0 (0.01) & 0.01 (0.03) & 0.01 (0.04) & 0.01 (0.01) & 0.02 (0.02) & 0.01 (0.04) & 0.92 (0.78) & 0 (0.02) & 0 (0.01) \\
\hline
SBC & 0 (0.01) & 0.01 (0.03) & 0.03 (0.05) & 0 (0.01) & 0.01 (0.02) & 0 (0.01) & 0.01 (0.02) & 0.02 (0.03) & 0 (0.01) & 0.01 (0.02) & 0.89 (0.76) & 0.01 (0.03) \\
\hline
SUB & 0.01 (0.02) & 0.01 (0.02) & 0.01 (0.02) & 0.01 (0.01) & 0 (0.01) & 0.01 (0.01) & 0 (0.01) & 0.01 (0.01) & 0 (0) & 0 (0) & 0 (0.01) & 0.93 (0.87) \\
\hline
\end{tabular}
\end{minipage}}
\end{table}

\newpage

 \begin{table}[h]
 \scriptsize
 \centering
 \caption{Offline  (real-time) within-group 2 instruction disassembly result} 
\begin{tabular}{|c|c|c|c|c|c|c|c|c|c|c|}
 \hline

{} & {ADIW} & {ANDI} & {CBR} & {CPI} & {LDI} & {ORI} & {SBCI} & {SBIW} & {SBR} & {SUBI} \\
\hline
ADIW & 0.91 (0.85) & 0 (0.01) & 0 (0.02) & 0.07 (0.1) & 0 (0) & 0 (0) & 0 (0) & 0.01 (0.01) & 0 (0) & 0 (0) \\
\hline
ANDI & 0.01 (0.01) & 0.88 (0.79) & 0.01 (0.02) & 0 (0.01) & 0.01 (0.03) & 0.01 (0.04) & 0.01 (0.02) & 0.03 (0.03) & 0.01 (0.01) & 0.04 (0.05) \\
\hline
CBR & 0.01 (0.01) & 0.03 (0.04) & 0.86 (0.78) & 0.01 (0.02) & 0.03 (0.04) & 0.01 (0.02) & 0.01 (0.01) & 0.02 (0.03) & 0 (0.02) & 0.01 (0.02) \\
\hline
CPI & 0.02 (0.06) & 0 (0.01) & 0.01 (0.01) & 0.94 (0.88) & 0 (0.01) & 0 (0.01) & 0.01 (0.01) & 0 (0.01) & 0.01 (0.01) & 0 (0) \\
\hline
LDI & 0.02 (0.02) & 0.01 (0.02) & 0.01 (0) & 0.01 (0.01) & 0.84 (0.76) & 0.02 (0.07) & 0.02 (0.03) & 0.03 (0.03) & 0.03 (0.03) & 0.01 (0.02) \\
\hline
ORI & 0 (0.01) & 0.02 (0.03) & 0.01 (0.02) & 0.01 (0.02) & 0.03 (0.02) & 0.86 (0.79) & 0.01 (0.03) & 0.01 (0.03) & 0.01 (0.02) & 0.01 (0.02) \\
\hline
SBCI & 0.01 (0.02) & 0.01 (0.01) & 0.01 (0.03) & 0 (0.01) & 0.01 (0.03) & 0 (0.01) & 0.92 (0.83) & 0.01 (0.02) & 0.01 (0.01) & 0.01 (0.03) \\
\hline
SBIW & 0 (0.01) & 0.01 (0.01) & 0.01 (0.02) & 0.02 (0.01) & 0.02 (0.01) & 0.01 (0.03) & 0.02 (0.03) & 0.88 (0.82) & 0.01 (0.03) & 0.02 (0.03) \\
\hline
SBR & 0.01 (0.01) & 0 (0.01) & 0 (0.01) & 0.01 (0.01) & 0 (0.02) & 0.01 (0.01) & 0.01 (0.04) & 0.02 (0.01) & 0.92 (0.86) & 0.02 (0.02) \\
\hline
SUBI & 0.01 (0.01) & 0.01 (0.01) & 0 (0.01) & 0.02 (0.03) & 0.01 (0.02) & 0.01 (0.03) & 0.01 (0.01) & 0.01 (0.01) & 0 (0.02) & 0.91 (0.86) \\
\hline

\end{tabular}
\end{table}

\newpage

\begin{table}[h]
 \centering
 \rotatebox{90}{
 \begin{minipage}{1.15\textwidth}
 \scriptsize
 \caption{Offline  (real-time) within-group 3 instruction disassembly result} 
\begin{tabular}
{|c|c|c|c|c|c|c|c|c|c|c|c|c|c|c|c|}
 \hline

{} & {ASR} & {CLR} & {COM} & {DEC} & {INC} & {LSL} & {LSR} & {NEG} & {ROL}& {ROR} & {SER} & {SWAP} & {TST}   \\
\hline
ASR & 0.85 (0.77) & 0.02 (0.02) & 0.02 (0.03) & 0.01 (0.04) & 0.01 (0.02) & 0 (0.01) & 0.01 (0.01) & 0.02 (0.03) & 0.01 (0.01) &0.01 (0.01) & 0.01 (0.01) & 0.01 (0.02) & 0.01 (0.02)  \\
\hline
CLR & 0.01 (0.04) & 0.85 (0.76) & 0.01 (0.02) & 0.01 (0.02) & 0.01 (0.01) & 0 (0.01) & 0.01 (0.01) & 0.02 (0.03) & 0.03 (0.02)& 0.01 (0.01) & 0.01 (0.01) & 0.01 (0.02) & 0.02 (0.03)   \\
\hline
COM & 0.01 (0.01) & 0.01 (0.02) & 0.89 (0.84) & 0.01 (0.01) & 0.01 (0.01) & 0 (0.01) & 0 (0.01) & 0 (0.01) & 0 (0)& 0 (0) & 0.01 (0.02) & 0.02 (0.01) & 0.04 (0.04)  \\
\hline
DEC & 0.01 (0.01) & 0.01 (0.02) & 0.01 (0.02) & 0.89 (0.82) & 0.01 (0.01) & 0.01 (0) & 0.01 (0.01) & 0.02 (0) & 0.01 (0.01)& 0.01 (0.03) & 0 (0.01) & 0.01 (0.02) & 0.01 (0.01)  \\
\hline
INC & 0.01 (0.01) & 0.01 (0.01) & 0.01 (0.02) & 0.01 (0.02) & 0.88 (0.8) & 0.01 (0.02) & 0.01 (0.01) & 0.01 (0.01) & 0.01 (0.01)& 0.02 (0.02) & 0.02 (0.01) & 0.02 (0.03) & 0.01 (0.03) \\
\hline
LSL & 0.01 (0.02) & 0.02 (0.02) & 0.02 (0.03) & 0.01 (0.02) & 0.01 (0.01) & 0.85 (0.77) & 0.01 (0.01) & 0.02 (0.03) & 0.01 (0.01)& 0.02 (0.02) & 0.02 (0.03) & 0.01 (0.02) & 0.01 (0.02) \\
\hline
LSR & 0 (0.03) & 0.01 (0.02) & 0.02 (0.03) & 0.01 (0.01) & 0.01 (0.01) & 0 (0.01) & 0.90 (0.75) & 0.01 (0.02) & 0.01 (0.02)& 0.02 (0.01) & 0.01 (0.04) & 0.01 (0.02) & 0 (0.03)\\
\hline
NEG & 0.01 (0.01) & 0.02 (0.03) & 0.02 (0.02) & 0.01 (0.01) & 0.01 (0.01) & 0.02 (0.01) & 0 (0.01) & 0.83 (0.8) & 0.01 (0.02) &0.01 (0.02) & 0.01 (0.02) & 0.02 (0.02) & 0.02 (0.02)  \\
\hline
ROL & 0.01 (0.01) & 0.03 (0.03) & 0.02 (0.01) & 0.01 (0.01) & 0.01 (0.02) & 0.03 (0.03) & 0.03 (0.03) & 0.03 (0.04) & 0.78 (0.7)&0.02 (0.03) & 0.02 (0.02) & 0.02 (0.03) & 0.01 (0.01)  \\
\hline
ROR & 0.01 (0.01) & 0.02 (0.02) & 0.01 (0.01) & 0.01 (0.03) & 0.01 (0.01) & 0.02 (0.02) & 0.01 (0.03) & 0.01 (0.02) & 0.02 (0.01)&0.84 (0.74) & 0.02 (0.05) & 0.01 (0.01) & 0.01 (0.02) \\
\hline
SER & 0.01 (0.01) & 0.01 (0.02) & 0.01 (0.03) & 0 (0.02) & 0.02 (0.02) & 0.01 (0.02) & 0.01 (0.02) & 0 (0.02) & 0.01 (0.01)& 0 (0.01) & 0.90 (0.79) & 0.01 (0.02) & 0.01 (0.01)  \\
\hline
SWAP & 0.02 (0.02) & 0.02 (0.03) & 0.04 (0.04) & 0.02 (0.03) & 0.01 (0.02) & 0.01 (0.03) & 0.01 (0.02) & 0.02 (0.03) & 0.01 (0.03)&0.01 (0.03) & 0.01 (0.02) & 0.80 (0.69) & 0.01 (0.02)  \\
\hline
TST & 0.01 (0.02) & 0.02 (0.02) & 0.01 (0.02) & 0.01 (0.02) & 0.02 (0.03) & 0.01 (0.01) & 0 (0.01) & 0 (0.01) & 0.01 (0.01)&0.01 (0.01) & 0.01 (0.01) & 0.01 (0.02) & 0.88 (0.8)  
\\ 
\hline
\end{tabular}
\end{minipage}}
\end{table}

\newpage

\begin{table}[h]
 \centering
 \rotatebox{90}{
 \begin{minipage}{1.15\textwidth}
 \scriptsize
 \caption{Offline  (real-time) within-group 4 instruction disassembly result} 
\begin{tabular}
{|p{0.6cm}|p{0.5cm}|p{0.5cm}|p{0.5cm}|p{0.5cm}|p{0.5cm}|p{0.5cm}|p{0.5cm}|p{0.5cm}|p{0.5cm}|p{0.5cm}|p{0.5cm}|p{0.5cm}|p{0.5cm}|p{0.5cm}|p{0.5cm}|p{0.5cm}|p{0.5cm}|p{0.5cm}|p{0.5cm}|p{0.5cm}|}
\hline
{} & {BRCC} & {BRCS} & {BREQ} & {BRGE} & {BRHC} & {BRHS} & {BRLO} & {BRLT} & {BRMI} & {BRNE}& {BRPL} & {BRSH} & {BRTC} & {BRTS} & {BRVC} & {BRVS} & {CALL} & {JMP} & {RCALL} & {RJMP}\\
\hline
BRCC & 0.81 (0.76) & 0 (0) & 0.01 (0.01) & 0.02 (0.02) & 0.01 (0.01) & 0.01 (0.02) & 0.02 (0.02) & 0.01 (0.01) & 0 (0) & 0.01 (0.02)& 0.02 (0.02) & 0.01 (0.01) & 0 (0.01) & 0.02 (0.04) & 0 (0.01) & 0.01 (0.01) & 0.02 (0.01) & 0.01 (0.01) & 0.01 (0.02) & 0.01 (0.02)  \\
\hline
BRCS & 0.07 (0.08) & 0.57 (0.46) & 0.01 (0.01) & 0.05 (0.04) & 0.01 (0.02) & 0.02 (0.03) & 0.04 (0.03) & 0.01 (0.02) & 0.04 (0.05) & 0.01 (0.02)& 0.01 (0.02) & 0.02 (0.01) & 0.01 (0.03) & 0.03 (0.02) & 0.01 (0.03) & 0.03 (0.03) & 0.02 (0) & 0.01 (0.01) & 0 (0.05) & 0.01 (0.02)  \\
\hline
BREQ & 0.02 (0.02) & 0 (0) & 0.80 (0.67) & 0.01 (0.01) & 0.02 (0.01) & 0 (0.01) & 0 (0.01) & 0.01 (0.01) & 0.01 (0.01) & 0.01 (0.01)&0.01 (0.01) & 0.01 (0.02) & 0.01 (0.04) & 0.02 (0.03) & 0.01 (0.02) & 0.01 (0.02) & 0.02 (0.02) & 0.01 (0.01) & 0.01 (0.01) & 0.01 (0.03)  \\
\hline
BRGE & 0.02 (0.04) & 0 (0) & 0.01 (0.02) & 0.76 (0.63) & 0 (0.01) & 0.02 (0.02) & 0.02 (0.03) & 0.01 (0) & 0.01 (0.02) & 0.01 (0.01)& 0.04 (0.02) & 0.01 (0.03) & 0.02 (0.02) & 0.01 (0.05) & 0 (0.01) & 0.02 (0.02) &0.01 (0.01) & 0.01 (0.02) & 0.02 (0.01) & 0.01 (0.02) \\
\hline
BRHC & 0.02 (0.03) & 0 (0) & 0.01 (0.02) & 0.02 (0.03) & 0.74 (0.6) & 0.01 (0.01) & 0 (0.01) & 0.02 (0.04) & 0.01 (0.03) & 0.01 (0.01)& 0.02 (0.03) & 0.01 (0.02) & 0.01 (0.01) & 0 (0.01) & 0.02 (0.04) & 0.03 (0.03) & 0.04 (0.04) & 0.01 (0.01) & 0.01 (0.01) & 0.03 (0.04)  \\
\hline
BRHS & 0.02 (0.05) & 0 (0) & 0.01 (0.02) & 0.02 (0.03) & 0.01 (0.01) & 0.80 (0.65) & 0 (0.02) & 0.02 (0.01) & 0.01 (0.01) & 0 (0.02)& 0.02 (0.03) & 0.02 (0.01) & 0.01 (0.02) & 0.02 (0.02) & 0.01 (0.02) & 0 (0.02) & 0.01 (0.01) & 0 (0.02) & 0 (0.02) & 0.01 (0.01) \\
\hline
BRLO & 0.01 (0.02) & 0 (0) & 0.01 (0.02) & 0.01 (0.02) & 0.01 (0.03) & 0.02 (0.01) & 0.80 (0.67) & 0.02 (0.03) & 0.01 (0.01) & 0 (0.01)& 0.02 (0.01) & 0.01 (0.03) & 0.01 (0.02) & 0.01 (0.01) & 0.01 (0.03) & 0.02 (0.01) & 0.01 (0.03) & 0 (0.01) & 0.01 (0) & 0.01 (0.01)\\
\hline
BRLT & 0.02 (0.01) & 0 (0) & 0.01 (0.01) & 0.01 (0.02) & 0 (0.02) & 0.01 (0.01) & 0.01 (0.01) & 0.82 (0.73) & 0.01 (0.01) & 0 (0.01)& 0.01 (0.01) & 0.02 (0.02) & 0.01 (0.02) & 0.01 (0.02) & 0.01 (0.02) & 0.01 (0.02) & 0.01 (0.02) & 0 (0.01) & 0.01 (0.01) & 0.02 (0.03)  \\
\hline
BRMI & 0.01 (0.01) & 0.01 (0.01) & 0.01 (0.01) & 0.01 (0.02) & 0.01 (0.02) & 0.02 (0.02) & 0.01 (0.01) & 0.02 (0.03) & 0.77 (0.66) & 0.01 (0.01)& 0.01 (0.01) & 0.03 (0.01) & 0.03 (0.03) & 0.01 (0.01) & 0 (0.02) & 0 (0.02) & 0.02 (0.02) & 0.01 (0.02) & 0.01 (0.03) & 0.02 (0.04)  \\

\hline
BRNE & 0.02 (0.01) & 0 (0) & 0.02 (0.02) & 0.02 (0.02) & 0.01 (0.01) & 0.01 (0.01) & 0.02 (0.02) & 0 (0.02) & 0.01 (0.01) & 0.80 (0.70)& 0.01 (0.01) & 0.02 (0.03) & 0 (0.01) & 0.01 (0.04) & 0.01 (0.01) & 0.01 (0.01) & 0.02 (0.02) & 0.01 (0.01) & 0.01 (0.02) & 0 (0) 
\\

\hline
BRPL & 0.03 (0.04) & 0 (0) & 0.03 (0.02) & 0.01 (0.01) & 0 (0.01) & 0.01 (0.02) & 0.02 (0.04) & 0.01 (0.01) & 0 (0) & 0.01 (0.01)& 0.78 (0.65) & 0.01 (0.03) & 0.03 (0.02) & 0.02 (0.02) & 0.01 (0.01) & 0.01 (0.01) & 0.01 (0.03) & 0 (0.02) & 0.01 (0.04) & 0 (0.04) \\

\hline
BRSH & 0 (0.02) & 0.01 (0.01) & 0.01 (0.01) & 0.01 (0.01) & 0.01 (0.02) & 0 (0) & 0.01 (0.02) & 0.01 (0.01) & 0.01 (0.01) & 0.01 (0.03)& 0.01 (0.03) & 0.81 (0.63) & 0.02 (0.03) & 0.01 (0.03) & 0.03 (0.03) & 0.01 (0.02) & 0.01 (0.04) & 0.02 (0.02) & 0.01 (0.01) & 0.01 (0.02) \\
\hline
BRTC & 0.01 (0.02) & 0 (0) & 0.01 (0.02) & 0.01 (0.03)  & 0.01 (0.02) & 0 (0.01) & 0.01 (0.01) & 0.03 (0.03) & 0.01 (0.01) & 0 (0.01)& 0.01 (0.03) &0.02 (0.03)   &0.81 (0.63) & 0.01 (0.03) & 0.03 (0.03) & 0.01 (0.02) & 0.01 (0.04) & 0.02 (0.02) & 0.01 (0.01) & 0.01 (0.02) \\

\hline
BRTS & 0.02 (0.04) & 0 (0.02) & 0 (0.02) & 0.02 (0.03) & 0.01 (0.02) & 0.01 (0.01) & 0.01 (0.03) & 0.01 (0.01) & 0.01 (0.01) & 0.01 (0.01)& 0.02 (0.02) & 0.01 (0.02) & 0.02 (0.01) & 0.79 (0.67) & 0 (0) & 0.02 (0.02) & 0.01 (0.02) & 0.01 (0.01) & 0.01 (0.02) & 0.01 (0.02) \\

\hline
BRVC & 0 (0) & 0.01 (0.01) & 0.01 (0.02) & 0 (0.01) & 0.01 (0.02) & 0.01 (0.02) & 0 (0.01) & 0.01 (0.01) & 0.01 (0.01) & 0.01 (0.01)& 0.00 (0.01) & 0.01 (0.01) & 0 (0.02) & 0.01 (0.01) & 0.87 (0.74) & 0.01 (0.01) & 0.02 (0.02) &0 (0.01)& 0 (0.01)& 0.01 (0.01) \\

\hline
BRVS & 0.01 (0.02) & 0 (0) & 0.01 (0.01) & 0.01 (0.01) & 0 (0.01) & 0.01 (0.02) & 0.02 (0.03) & 0 (0.01) & 0.01 (0.01) & 0.01 (0.02) & 0.02 (0.02) & 0.01 (0.01) & 0 (0) & 0.02 (0.02) & 0 (0.01) & 0.84 (0.74) & 0.01 (0.02) & 0.02 (0.01) & 0.01 (0.02) & 0 (0.01)\\

\hline
CALL & 0.01 (0.01) & 0.01 (0) & 0.02 (0.03) & 0 (0.02) & 0.01 (0.01) & 0.02 (0.01) & 0.02 (0.03) & 0.02 (0.01) & 0.01 (0.02) & 0.02 (0.04)& 0 (0.02) & 0.01 (0.02) & 0.01 (0.02) & 0.01 (0.02) & 0.01 (0.03) & 0.01 (0.01) & 0.79 (0.63) & 0.01 (0.02) & 0.01 (0.01) & 0.01 (0.04) \\

\hline
JMP & 0.02 (0.02) & 0 (0) & 0.01 (0.02) & 0.02 (0.05) & 0.02 (0.02) & 0.01 (0.02) & 0.01 (0.02) & 0.03 (0.03) & 0.02 (0.02) & 0 (0.02) & 0.01 (0.01) & 0.01 (0.02) & 0.01 (0.02) & 0.02 (0.04) & 0.01 (0.01) & 0.01 (0.03) & 0.01 (0.03) & 0.76 (0.57) & 0.01 (0.02) & 0.01 (0.01)  \\

\hline
RCALL & 0.02 (0.03) & 0 (0) & 0 (0.01) & 0.02 (0.03) & 0.01 (0.01) & 0.01 (0.01) & 0 (0.01) & 0.01 (0.01) & 0 (0) & 0.01 (0.01) & 0.01 (0.01) & 0.01 (0.03) & 0.02 (0.01) & 0.02 (0.04) & 0.01 (0.01) & 0.03 (0.05) & 0.01 (0.01) & 0.01 (0.01) & 0.80 (0.69) & 0 (0.01)
\\
\hline
RJMP & 0.01 (0.01) & 0 (0.01) & 0.01 (0.02) & 0 (0.01) & 0.01 (0.01) & 0 (0) & 0.01 (0.03) & 0.03 (0.04) & 0.01 (0.02) & 0.01 (0.01) & 0 (0.01) & 0.01 (0.02) & 0.01 (0.01) & 0 (0.01) & 0.01 (0.01) & 0 (0) & 0.02 (0.02) & 0.01 (0.02) & 0 (0) & 0.85 (0.75)  \\

\hline
\end{tabular}
\end{minipage}
}
\end{table}

\newpage

 \begin{table}[h]
 \scriptsize
 \centering
 \caption{Offline  (real-time) within-group 5 instruction disassembly result} 
\begin{tabular}{|c|c|c|c|}
 \hline
{} & {LD} & {LDD} & {LDS}  
\\ \hline
LD & 0.98 (0.95)  & 0.01 (0.03)  &0.01 (0.02) 
\\ \hline
LDD & 0.01 (0.02)  & 0.97 (0.91)  &0.02 (0.07)  
\\ \hline
LDS & 0.02 (0.02)  &0.02 (0.06)   & 0.97 (0.92) 
\\ \hline
\end{tabular}
\end{table}

 \begin{table}[h]
 
 \scriptsize
 \centering
 \caption{Offline (real-time) within-group 6 instruction disassembly result} 
\begin{tabular}{|c|c|c|}
 \hline
{} & {LPM} & {ELPM}  
\\ \hline
LPM & 0.97 (0.94)  & 0.03 (0.06) 
\\ \hline
ELPM & 0.03 (0.05)  & 0.97 (0.95) 
\\ \hline
\end{tabular}
\end{table}

\afterpage{
\begin{table}[h]
 \centering
 \rotatebox{90}{
 \begin{minipage}{1.2\textwidth}
 \scriptsize
 \caption{Offline  (real-time) within-group 7 instruction disassembly result} 
\begin{tabular}{|c|c|c|c|c|c|c|c|c|c|c|c|c|c|c|}
\hline
{} & {CLC} & {CLN} & {CLS} & {CLT} & {CLV} & {CLZ} & {SEC} & {SEH} & {SEI} & {SEN} & {SES} & {SET} & {SEV} & {SEZ} \\
\hline
CLC & 0.88 (0.66) & 0.01 (0.02) & 0 (0.02) & 0.01 (0.02) & 0.01 (0.04) & 0.01 (0.03) & 0.01 (0.02) & 0.01 (0.04) & 0.02 (0.04) & 0 (0.01) & 0.01 (0.03) & 0.02 (0.02) & 0.01 (0.02) & 0.01 (0.03) \\

\hline
CLN & 0.01 (0.02) & 0.88 (0.72) & 0.01 (0.01) & 0 (0.02) & 0.02 (0.04) & 0.02 (0.02) & 0.01 (0.04) & 0 (0.02) & 0.01 (0.02) & 0 (0.01) & 0.01 (0.02) & 0.01 (0.02) & 0.01 (0.02) & 0.01 (0.02) \\

\hline
CLS & 0 (0.01) & 0.01 (0.02) & 0.86 (0.71) & 0.01 (0.02) & 0.01 (0.02) & 0.03 (0.05) & 0.02 (0.04) & 0.01 (0.01) & 0.02 (0.02) & 0 (0.03) & 0.01 (0.01) & 0 (0.01) & 0.01 (0.02) & 0.05 \\

\hline
CLT & 0.01 (0.02) & 0.02 (0.03) & 0.01 (0.02) & 0.87 (0.74) & 0.01 (0.01) & 0.02 (0.03) & 0.01 (0.03) & 0.01 (0.01) & 0.01 (0.02) & 0.02 (0.03) & 0.01 (0.03) & 0.01 (0.01) & 0.01 (0.01) & 0 (0.01) \\

\hline
CLV & 0.01 (0.03) & 0.01 (0.03) & 0.01 (0.02) & 0.01 (0.03) & 0.89 (0.72) & 0.01 (0.02) & 0.01 (0.02) & 0.01 (0.01) & 0.01 (0.03) & 0.01 (0.02) & 0 (0.01) & 0.01 (0.02) & 0.01 (0.02) & 0.01 (0.03) \\

\hline
CLZ & 0 (0.02) & 0.01 (0.02) & 0.01 (0.03) & 0.01 (0.02) & 0.01 (0.02) & 0.87 (0.68) & 0.01 (0.04) & 0.01 (0.02) & 0.01 (0.03) & 0.01 (0.04) & 0.01 (0.03) & 0.01 (0.01) & 0.01 (0.02) & 0.01 (0.02) \\

\hline
SEC & 0 (0.02) & 0.01 (0.01) & 0.01 (0.02) & 0.02 (0.03) & 0 (0.03) & 0.01 (0.02) & 0.89 (0.72) & 0.01 (0.01) & 0 (0.02) & 0.01 (0.03) & 0.01 (0.01) & 0.01 (0.03) & 0.02 (0.03) & 0 (0.02) \\

\hline
SEH & 0.01 (0.03) & 0.01 (0.02) & 0.04 (0.07) & 0 (0.03) & 0.01 (0.02) & 0.02 (0.02) & 0.01 (0.03) & 0.84 (0.65) & 0.01 (0.01) & 0.01 (0.02) & 0.01 (0.03) & 0.01 (0.03) & 0.01 (0.02) & 0.01 (0.04) \\

\hline
SEI & 0.01 (0.04) & 0.01 (0.03) & 0 (0.01) & 0 (0.02) & 0.01 (0.03) & 0.02 (0.03) & 0.01 (0.02) & 0 (0.02) & 0.87 (0.69) & 0.01 (0.01) & 0.02 (0.02) & 0 (0.01) & 0.01 (0.02) & 0.02 (0.03) \\

\hline
SEN & 0 (0.01) & 0 (0.01) & 0 (0.02) & 0.01 (0.02) & 0.01 (0.02) & 0.02 (0.03) & 0 (0.02) & 0.01 (0.01) & 0.02 (0.03) & 0.90 (0.79) & 0.01 (0.02) & 0.01 (0.02) & 0 (0.01) & 0.01 (0.01)
\\

\hline
SES & 0.01 (0.02) & 0.01 (0.02) & 0.01 (0.02) & 0 (0.02) & 0 (0.01) & 0.01 (0.02) & 0.01 (0.02) & 0 (0.02) & 0.01 (0.03) & 0.01 (0.02) & 0.91 (0.76) & 0 (0.01) & 0.01 (0.02) & 0 (0.02)
\\
\hline
SET & 0.01 (0.02) & 0.01 (0.02) & 0 (0) & 0.02 (0.05) & 0.01 (0.02) & 0.01 (0.03) & 0.01 (0.04) & 0.01 (0.04) & 0.02 (0.03) & 0.01 (0.02) & 0.01 (0.01) & 0.87 (0.69) & 0.01 (0.03) & 0.01 (0.02) \\

\hline
SEV & 0.01 (0.02) & 0.01 (0.02) & 0.01 (0.02) & 0 (0.02) & 0.01 (0.03) & 0 (0.02) & 0.01 (0.04) & 0.01 (0.02) & 0.01 (0.03) & 0.02 (0.02) & 0.01 (0.03) & 0 (0.02) & 0.90 (0.71) & 0.01 (0.02) \\

\hline
SEZ & 0.01 (0.02) & 0.01 (0.02) & 0.01 (0.02) & 0 (0.02) & 0.02 (0.04) & 0.01 (0.02) & 0.02 (0.03) & 0.02 (0.02) & 0.01 (0.02) & 0.01 (0.02) & 0.01 (0.02) & 0.01 (0.03) & 0.01 (0.03) & 0.86 (0.70) \\

\hline

\end{tabular}
\end{minipage}}
\end{table}
}

\newpage

\afterpage{
\begin{table}[h]
 \centering
 \rotatebox{90}{
 \begin{minipage}{1.15\textwidth}
 \scriptsize
 \caption{Offline  (real-time) within-group 8 instruction disassembly result} 
\begin{tabular}{|c|c|c|c|c|c|c|c|c|c|c|c|c|c|c|}

\hline
{} & {BCLR} & {BLD} & {BRBC} & {BRBS} & {BSET} & {BST} & {CBI} & {SBI} & {SBIC} & {SBIS} & {SBRC} & {SBRS} \\
\hline
BCLR & 0.80 (0.61) & 0.02 (0.03) & 0.03 (0.05) & 0.01 (0.02) & 0.02 (0.04) & 0.01 (0.05) & 0.04 (0.04) & 0.01 (0.01) & 0.03 (0.06) & 0.02 (0.04) & 0.01 (0.02) & 0.02 (0.02) \\

\hline
BLD & 0.01 (0.02) & 0.90 (0.75) & 0.02 (0.04) & 0.01 (0.02) & 0.01 (0.02) & 0.01 (0.04) & 0 (0) & 0 (0) & 0.01 (0.03) & 0.01 (0.01) & 0 (0.03) & 0.02 (0.03) \\

\hline
BRBC & 0.01 (0.02) & 0.01 (0.02) & 0.94 (0.88) & 0.01 (0.01) & 0 (0.01) & 0 (0.01) & 0 (0.01) & 0 (0.01) & 0 & 0.01 (0.01) & 0.01 (0.01) & 0 (0.01) \\

\hline
BRBS & 0 (0.01) & 0.01 (0.01) & 0.01 (0.03) & 0.93 (0.79) & 0 (0.01) & 0 (0) & 0 (0) & 0 (0) & 0.01 (0.05) & 0.01 (0.04) & 0 (0.03) & 0.01 (0.02) \\

\hline
BSET & 0.01 (0.01) & 0.02 (0.02) & 0 (0.01) & 0 (0.01) & 0.90 (0.77) & 0.01 (0.05) & 0.01 (0.01) & 0 (0) & 0.01 (0.03) & 0.01 (0.03) & 0.02 (0.04) & 0.01 (0.01) \\

\hline
BST & 0.02 (0.02) & 0.01 (0.01) & 0 (0) & 0 (0.01) & 0.01 (0.02) & 0.92 (0.86) & 0 (0) & 0 (0) & 0.01 (0.02) & 0.01 (0.02) & 0.01 (0.02) & 0 (0.02) \\

\hline
CBI & 0.01 (0.02) & 0.01 (0) & 0.02 (0.03) & 0.01 (0.02) & 0 (0) & 0 (0) & 0.85 (0.72) & 0.08 (0.19) & 0.01 (0.01) & 0 (0) & 0.01 (0) & 0 (0) \\

\hline
SBI & 0.01 (0.01) & 0 (0) & 0 (0.02) & 0 (0) & 0 (0) & 0 (0) & 0.03 (0.04) & 0.95 (0.03) & 0 (0)& 0 (0)& 0 (0)& 0 (0)  \\

\hline
SBIC & 0.01 (0.02) & 0.01 (0.02) & 0 (0.01) & 0.01 (0.03) & 0.01 (0.02) & 0 (0.02) & 0 (0) & 0 (0) & 0.92 (0.80) & 0.01 (0.03) & 0.01 (0.03) & 0.02 (0.03) \\

\hline
SBIS & 0 (0.02) & 0.01 (0.01) & 0.01 (0.01) & 0.01 (0.02) & 0.01 (0.02) & 0.01 (0.02) & 0 (0) & 0 (0) & 0.01 (0.05) & 0.91 (0.78) & 0.02 (0.04) & 0.02 (0.03) \\
\hline
SBRC & 0.01 (0.02) & 0 (0.02) & 0 (0) & 0.02 (0.03) & 0.01 (0.01) & 0 (0.01) & 0 (0) & 0 (0) & 0.02 (0.03) & 0.01 (0.03) & 0.92 (0.79) & 0 (0.05) \\

\hline
SBRS & 0 (0.01) & 0.01 (0.01) & 0 (0) & 0.01 (0.02) & 0.01 (0.01) & 0 (0.01) & 0 (0) & 0 (0) & 0.01 (0.02) & 0 (0.01) & 0.03 (0.06) & 0.92 (0.83) \\

\hline
\end{tabular}
\end{minipage}}
\end{table}
}

\end{document}